\documentclass[11pt]{article}
\usepackage{epstopdf}
\usepackage{color}
\usepackage{subfigure}
\usepackage{amsmath}
\usepackage{amssymb}
\usepackage{graphicx,color}
\usepackage{cite}
\usepackage{enumerate}
\usepackage{amsthm}
\usepackage{amsfonts,mathrsfs}
\usepackage{geometry}
\usepackage{ifthen}
\usepackage{graphicx}
\usepackage{float}
\usepackage{bm}
\usepackage[caption=false]{subfig}
\usepackage[utf8]{inputenc}

\begin{document}

\title{\bf Uncertainty of quantum channels via modified generalized variance and modified generalized Wigner-Yanase-Dyson skew information}

\vskip0.1in
\author{\small Cong Xu$^1$, Zhaoqi Wu$^1$\thanks{Corresponding author. E-mail: wuzhaoqi\_conquer@163.com}, Shao-Ming
Fei$^{2,3}$\\
{\small\it  1. Department of Mathematics, Nanchang University,
Nanchang 330031, China}\\
{\small\it  2. School of Mathematical Sciences, Capital Normal University, Beijing 100048, China}\\
{\small\it  3. Max-Planck-Institute for Mathematics in the Sciences,
04103 Leipzig, Germany}}
\date{}
\maketitle

\noindent {\bf Abstract} {\small }\\
Uncertainty relation is a fundamental issue in quantum mechanics and
quantum information theory. By using modified generalized variance
(MGV), and modified generalized Wigner-Yanase-Dyson skew information
(MGWYD), we identify the total and quantum uncertainty of quantum
channels. The elegant properties of the total uncertainty of quantum
channels are explored in detail. In addition, we present a trade-off
relation between the total uncertainty of quantum channels and the
entanglement fidelity, and establish the relationships between the
total uncertainty and entropy exchange/coherent information.
Detailed examples are given to the explicit formulas of the total
uncertainty and the quantum uncertainty of quantum channels.
Moreover, utilizing a realizable experimental measurement scheme by
using the Mach-Zehnder interferometer proposed in \cite{NGS}, we
discuss how to measure the total/quantum uncertainty of quantum
channels for pure states.

\vskip 0.1 in


\vskip0.2 in

\noindent {\bf 1 Introduction}\\\hspace*{\fill}\\
The quantum uncertainty is closely related to quantum measurements.
As an extremely important issue in the quantum physics, the
uncertainty relation has been initially put
forward by Heisenberg \cite{HEISENBERG} and Robertson \cite{ROBERT}.
Uncertainty is usually quantified by variance or entropies. The
total uncertainty of the observable in quantum states is quantified
by variance, which is a mixture of classical uncertainty and quantum
uncertainty \cite{LUO1,LUO2,LUO3}. Recently, Gudder \cite{GUDDER}
has introduced the modified variance of arbitrary operator (not
necessarily Hermitian). In addition, Dou and Du \cite{DD1,DD2} have
proposed a Heisenberg-type uncertainty relation and a
Schr\"{o}dinger-type uncertainty relation based on the modified
variance. Sun and Li \cite{SL} have proposed the total uncertainty of quantum
channels in terms of the modified variance.

The quantum uncertainty can be also described by skew information.
The skew information has been originally proposed by Wigner and
Yanase \cite{WY}, which is termed as {\it Wigner-Yanase} (WY) skew
information. While a more general quantity has been suggested
by Dyson, which is now called the {\it Wigner-Yanase-Dyson} (WYD)
skew information \cite{WY}. This quantity has been further
generalized in \cite{CL} and termed as {\it generalized
Wigner-Yanase-Dyson} (GWYD) skew information. The relation among the
WYD skew information, GWYD skew information and uncertainty
relations have been studied extensively \cite{YK1,YK2,CAIL}.
As is well known, the observables and Hamiltonians
in quantum mechanics are assumed to be Hermitian operators
mathematically. The framework of non-Hermitian quantum mechanics,
however, has also been taken into account and has attracted much
attention \cite{MN}. Besides, many operators such as quantum gates
\cite{NC}, generalized quantum gates \cite{LGL} and the Kraus
operators of a quantum channel \cite{NC} are not necessarily
Hermitian. Therefore, it is desirable to introduce the corresponding
definitions of the above-mentioned skew information for
pseudo-Hermitian and/or PT-symmetric quantum mechanics
\cite{BBS,MKGE,GSDM,RMER,CJHY,TWYH}. By considering an arbitrary
operator which may be non-Hermitian, Dou and Du \cite{DD2} have
introduced the {\it modified Wigner-Yanase-Dyson} (MWYD) skew
information, while Wu, Zhang and Fei \cite{WZWLF,WZFL} have further
introduced the {\it modified generalized Wigner-Yanase-Dyson}
(MGWYD) skew information.

Quantum channels characterize the general evolutions of quantum
systems \cite{BG,NC}, which include the quantum measurements as
special cases. The past few years have witnessed a great deal of
researches on quantum channels
\cite{AD,CQYP,FF,YFMF,MS,ZHA,HSTR,MR1,MK,LZZ,MR2,MM,TFIY,WILDE}. The
interaction between a quantum system and the external environment
would lead to information loss and disturbance on the quantum
states. Until now, many kinds of formulations have been established
on the trade-off relations between the information and the
disturbance \cite{BGD,BFI,SST,Luo4,KJR,MAS,KJJY,SP,SKU,PS}.
Schumacher \cite{SB1} has introduced the entanglement fidelity which
provides a measure of how well the entanglement between a quantum
system and an auxiliary system is preserved under the quantum
channel.


The main purpose of this paper is to provide a
quantification of the total as well as the quantum uncertainties of quantum
channels in terms of the modified generalized variance (MGV) and MGWYD skew information, respectively.
We give an extension of the trade-off relation between the total uncertainty of
quantum channels and the entanglement fidelity. By recalling two
important information quantities, the entropy exchange \cite{SB1} and
coherent information \cite{SB2}, we also generalize the relations
between the total uncertainty of quantum channels and the entropy
exchange/coherent information introduced in \cite{SL} to a more general case.

The rest of the paper is formulated as follows. In Section 2, we
introduce the total uncertainty of quantum channels based on MGV and
prove that it satisfies several fundamental properties. In Section
3, we generalize a trade-off relation between the total uncertainty
of quantum channels and the entanglement fidelity. Furthermore, by
using the general trade-off relation, we also explore the
relationship between the total uncertainty of quantum channels and
the entropy exchange/coherent information. In Section 4, we
calculate the total and quantum uncertainty of some typical quantum
channels, and derive the explicit formulas of the uncertainties of
the quantum channel in Example 5 for two special classes of states,
the Werner states and isotropic states. We also illustrate the
trade-off relation (18), inequalities (19) and (20) by using Example
6. We use an experimentally feasible protocol given in \cite{NGS} to
measure the total/quantum uncertainty of quantum channels for pure
states. Some concluding remarks are given in Section 5.

\vskip0.1in

\noindent {\bf 2 The total uncertainty of quantum channels based on {MGV}}\\\hspace*{\fill}\\

Let $\mathcal{H}$ be a $d$-dimensional Hilbert space. Denote by
$\mathcal{B(H)}$, $\mathcal{S(H)}$ and $\mathcal{D(H)}$ the set of
all bounded linear operators, Hermitian operators and density
operators on $\mathcal{H}$, respectively. For $\rho\in
\mathcal{D(H)}$ and $K\in \mathcal{B(H)}$, the {\it modified
generalized variance} (MGV) of the bounded linear operator ${K}$ in
$\rho$ is defined by \cite{WZWLF},
 \begin{align}\label{eq1}
  V^{\alpha,\beta}({\rho},K)
  =&\frac{1}{2}(\mathrm{tr}\rho K_0^\dag K_0+\mathrm{tr}\rho^{\alpha+\beta}K_0\rho^{1-\alpha-\beta}K_0^\dag) \notag \\
  =&\frac{1}{2}(\mathrm{tr}\rho K^\dag K+\mathrm{tr}\rho^{\alpha+\beta}K\rho^{1-\alpha-\beta}K^\dag)-{|\mathrm{tr}\rho K|}^{2},~~~\alpha,\beta\geq
0,~\alpha+\beta\leq 1
\end{align}with $K_0=K-\mathrm{tr}\rho K$. Note that the
modified generalized variance $V^{\alpha,\beta}({\rho},K)$ is
non-negative and $V^{\alpha,\beta}({\rho},K)$ is concave in $\rho$.
Analogizing the idea in Ref.\cite{LUO2},
$V^{\alpha,\beta}({\rho},K)$ can be also split into quantum and
classical parts as \cite{WZWLF},
  \begin{equation}\label{eq2}
  V^{\alpha,\beta}({\rho},K)=Q^{\alpha,\beta}({\rho},K)+C^{\alpha,\beta}({\rho},K),
  ~~~\alpha,\beta\geq 0,~\alpha+\beta\leq 1,
  \end{equation}
where
\begin{align}\label{eq3}
  Q^{\alpha,\beta}(\rho,K)
  =&\frac{1}{2}\mathrm{tr}([\rho^\alpha,K_0]^\dag[\rho^\beta,K_0]\rho^{1-\alpha-\beta})\notag \\
  =&\frac{1}{2}(\mathrm{tr}\rho K_0^\dag K_0-\mathrm{tr}\rho^{\alpha}K_0\rho^{1-\alpha}K_0^\dag-\mathrm{tr}\rho^{\beta}K_0\rho^{1-\beta}K_0^\dag+\mathrm{tr}\rho^{\alpha+\beta}K_0\rho^{1-\alpha-\beta}K_0^\dag)\notag \\
  =&\frac{1}{2}(\mathrm{tr}\rho K^\dag K -\mathrm{tr}\rho^{\alpha}K\rho^{1-\alpha}K^\dag-\mathrm{tr}\rho^{\beta}K\rho^{1-\beta}K^\dag+\mathrm{tr}\rho^{\alpha+\beta}K\rho^{1-\alpha-\beta}K^\dag),
  \end{align}
  where $\alpha,\beta\geq 0,~\alpha+\beta\leq 1$,
which is called {\it modified generalized Wigner-Yanase-Dyson}
(MGWYD) skew information quantifying the quantum
uncertainty of ${K}$ in $\rho$, and
 \begin{align}\label{eq4}
    C^{\alpha,\beta}({\rho},K)
  =&V^{\alpha,\beta}({\rho},K)-Q^{\alpha,\beta}({\rho},K) \notag \\
  =&\frac{1}{2}(\mathrm{tr}\rho^{\alpha}K_0\rho^{1-\alpha}K_0^\dag+\mathrm{tr}\rho^{\beta}K_0\rho^{1-\beta}K_0^\dag) \notag \\
  =&\frac{1}{2}(\mathrm{tr}\rho^{\alpha}K\rho^{1-\alpha}K^\dag+\mathrm{tr}\rho^{\beta}K\rho^{1-\beta}K^\dag)-{|\mathrm{tr}\rho K|}^{2}
  ,~~~\alpha,\beta\geq 0,~\alpha+\beta\leq 1,
 \end{align}
which quantifies the classical uncertainty of ${K}$ in $\rho$.

The WY skew information \cite{WY}
$Q(\rho,H)=-\frac{1}{2}\mathrm{tr}[\sqrt{\rho},H]^2$ has been
introduced as a measure of information content of observables not
commuting with (skew to) the conserved observable $H$ in the state
$\rho$. However, as we have mentioned, some important operators are
non-Hermitian. With respect to the WY skew information,
$Q(\rho,H)=-\frac{1}{2}\mathrm{tr}[\sqrt{\rho},H^\dag][\sqrt{\rho},H]$,
the modified Wigner-Yanase skew information
$Q(\rho,K)=-\frac{1}{2}\mathrm{tr}[\sqrt{\rho},\\K^\dag][\sqrt{\rho},K]$
has been defined in \cite{DD1,SLYS} (up to a constant factor in
\cite{SLYS}), which may be interpreted as a measure of information
content of observables skew to the Hamiltonian $K$ in the state
$\rho$ in PT-symmetric quantum mechanics. In this regard, we can say
that the generalized quantity $Q^{\alpha,\beta}(\rho,K)$ provides a
family of quantifiers of such information content.

It is proved that \cite{CAIL}
$Q^{\alpha,\beta}(\rho,H)=-\frac{1}{2}\mathrm{tr}([\rho^\alpha,H]
[\rho^\beta,H]\rho^{1-\alpha-\beta})$ is convex in $\rho$ when
$\alpha,\beta\in[0,1]$ with $\alpha+2\beta\leq1$ and
$2\alpha+\beta\leq1$.
By using the Morozova-Chentsov function of a regular metric, the WYD
skew information has been extended to the metric adjusted skew
information (of a state with respect to a conserved observable)
\cite{HF}, which is a non-negative quantity bounded by the variance
(of an observable in a state) that vanishes for observables
commuting with the state. Note that $2Q^{\alpha,\beta}(\rho,H)$ is a
metric adjusted skew information \cite{CAIL}, where the
Morozova-Chentsov function is
\begin{align*}
c(x,y)
=&\frac{1}{(x-y)^2}[(x^\alpha-y^\alpha)(x^{1-\alpha}-y^{1-\alpha})+(x^\beta-y^\beta)(x^{1-\beta}-y^{1-\beta})\\
-&(x^{\alpha+\beta}-y^{\alpha+\beta})(x^{1-\alpha-\beta}-y^{1-\alpha-\beta})].
\end{align*}
The MGWYD skew information $Q^{\alpha,\beta}(\rho,K)$ is also convex
in $\rho$ when $\alpha,\beta\in[0,1]$ with $\alpha+2\beta\leq1$ and
$2\alpha+\beta\leq1$ \cite{WZFL}. By using the Lieb's theorem
\cite{NC,EHL}, it can be easily obtained that
$C^{\alpha,\beta}({\rho},K)$ is concave in $\rho$ for
$\alpha,\beta\geq 0,~\alpha+\beta\leq 1$. These facts demonstrate
that $Q^{\alpha,\beta}(\rho,K)$ and $C^{\alpha,\beta}({\rho},K)$
could serve as the quantifiers of quantum and classical uncertainty,
respectively.

\vskip0.1in

Consider a quantum channel $\Phi$ given by Kraus operators $\{K_i\}$, $\Phi(\rho)=\sum_{i}K_i\rho K_i^\dag$.
We introduce the total uncertainty of the quantum channel $\Phi$ based on MGV,
   \begin{equation}\label{eq5}
   V^{\alpha,\beta}({\rho},\Phi)=\sum_{i}V^{\alpha,\beta}({\rho},K_i), ~~~\alpha,\beta\geq 0,~\alpha+\beta\leq 1.
   \end{equation}
By using Eq.
(\ref{eq1}), $V^{\alpha,\beta}({\rho},\Phi)$ can be further
rewritten as,
\begin{align}\label{eq6}
V^{\alpha,\beta}({\rho},\Phi)
=&\frac{1+\sum_{i}\mathrm{tr}\rho^{\alpha+\beta}K_i \rho^
{1-\alpha-\beta} K_i^\dag}{2}-\sum_{i}{|\mathrm{tr}\rho K_i|}^{2}
\notag \\
=&\frac{1+\mathrm{tr}\rho^{\alpha+\beta}\Phi(\rho^{1-\alpha-\beta})
}{2}-\sum_{i}{|\mathrm{tr}\rho K_i|}^{2}    ,~~~\alpha,\beta\geq
0,~\alpha+\beta\leq1.
\end{align}When $\alpha+\beta=1$,
$V^{\alpha,\beta}(\rho,\Phi)$ reduces to $V(\rho,\Phi)$, and Eq. (\ref{eq6}) reduces to the following equation in \cite{SL}, i.e.,
\begin{align}\label{eq7}
V({\rho},\Phi)=\frac{1+\mathrm{tr}\rho\Phi(\mathbf{1})}{2}-\sum_{i}{|\mathrm{tr}\rho K_i|}^{2},
\end{align} where $\mathbf{1}$ is the identity operator and $\Phi(\mathbf{1})=\sum_iK_iK_i^\dag$.

Let $\{E_i\}$ and $\{F_j\}$ be two sets of Kraus operators of the
quantum channel $\Phi$. Then there exists a unitary matrix
${U=(u_{ij})}$ such that $E_i=\sum_j u_{ij} F_j$ for any $i$
\cite{NC}. From (\ref{eq6}) we can easily check that
$V^{\alpha,\beta}({\rho},\Phi)$ is independent of the choice of the
Kraus operators of $\Phi$, namely, $V^{\alpha,\beta}({\rho},\Phi)$
is well-defined. For each $\alpha,\beta$ with $\alpha,\beta\geq
0,~\alpha+\beta\leq 1$, we can prove that the uncertainty
$V^{\alpha,\beta}({\rho},\Phi)$ has the following elegant
properties:

(i) (Non-negativity) $V^{\alpha,\beta}({\rho},\Phi)\geq 0$, with the
equality holds if and only if $\rho^
\frac{\alpha+\beta}{2}K_i\rho^\frac{1-\alpha-\beta}{2}\\
=K_i\sqrt{\rho}=(\mathrm{tr}\rho K_i)\sqrt{\rho}$ for any $i$.

(ii) (Linearity) $V^{\alpha,\beta}({\rho},\Phi)$ is
positive-real-linear with respect to the channel $\Phi$, i.e.,
$V^{\alpha,\beta}({\rho},\lambda_1\Phi_1+\lambda_2\Phi_2)=\lambda_1
V^{\alpha,\beta}({\rho},\Phi_1)+\lambda_2
V^{\alpha,\beta}({\rho},\Phi_2)$ for any $\lambda_1,\lambda_2 \geq
0$ and any quantum channels $\Phi_1$ and $\Phi_2$.

(iii) (Concavity) $V^{\alpha,\beta}(\rho,\Phi)$ is concave with
respect to $\rho$, i.e.,
$V^{\alpha,\beta}({\sum_j\lambda_j\rho_j},\Phi)\geq
\sum_j\lambda_jV^{\alpha,\beta}(\rho_j,\Phi)$, where $\lambda_j\geq
0$ for each $j$ with $\sum_j\lambda_j=1$.

(iv) (Unitary invariance) $V^{\alpha,\beta}({U\rho U^\dag},U\Phi
U^\dag)=V^{\alpha,\beta}({\rho},\Phi)$ for any unitary operators
$U$, where $U\Phi U^\dag(\rho)=\sum_i(U K_iU^\dag)$ $\rho (U K_i
U^\dag)^\dag$ with $\Phi(\rho)=\sum_{i}K_i\rho K_i^\dag$.

(v) (Ancillary independence)
$V^{\alpha,\beta}(\rho^{A}\otimes\rho^{B},\Phi^A\otimes\mathcal{I}^B)=V^{\alpha,\beta}(\rho^A,\Phi^A)$,
where $\rho^{A}$ and $\rho^{B}$ are any states of systems $A$ and
$B$, respectively, and $\mathcal{I}^B$ is the identity channel on
system $B$.

(vi) (Increasing under partial trace)
$V^{\alpha,\beta}(\rho^{AB},\Phi^A\otimes\mathcal{I}^B)\leq
V^{\alpha,\beta}(\rho^A,\Phi^A)$, where $\rho^{AB}$ is any bipartite
state on joint system $AB$, ${(\rho^A)}^{\alpha+\beta}=\mathrm{tr}_B
[(\rho^{AB})^{\alpha+\beta}]$ is the reduced state on system $A$,
and $\mathcal{I}^B$ is the identity channel on system $B$.

(vii) (Subadditivity)
$V^{\alpha,\beta}(\rho^{AB},\Phi^A\otimes\mathcal{I}^B+\mathcal{I}^A\otimes\Phi^B)\leq
V^{\alpha,\beta}(\rho^A,\Phi^A)+V^{\alpha,\beta}(\rho^B,\Phi^B)$,
where ${(\rho^A)}^{\alpha+\beta}=\mathrm{tr}_B
[(\rho^{AB})^{\alpha+\beta}]$ and
$(\rho^B)^{\alpha+\beta}=\mathrm{tr}_A[(\rho^ {AB})^{\alpha+\beta}]$
are the reduced states with respect to the subsystems $A$ and $B$,
$\Phi^A$ and $\Phi^B$ are the channels on systems $A$ and $B$, and
$\mathcal{I}^A$ and $\mathcal{I}^B$ are the identity channels on
systems $A$ and $B$, respectively.

\vskip0.1in

The above properties can be proved in the following way. From the
definition, $V^{\alpha,\beta}({\rho},\Phi)\geq 0$ is obvious.
$V^{\alpha,\beta}({\rho},\Phi)= 0$ if and
   only if $V^{\alpha,\beta}({\rho},K_i)=0$ for any $i$, i.e.,
   $\mathrm{tr}(\rho^{\alpha+\beta} K_{i0}\rho^{1-\alpha-\beta} K_{i0}^\dag)=\mathrm{tr}(\rho K_{i0}^\dag K_{i0})=0$, which is equivalent to $\rho^ \frac{\alpha+\beta}{2} K_i\rho^\frac{1-\alpha-\beta}{2}=K_i \sqrt{\rho}=(\mathrm{tr} \rho K_i)\sqrt{\rho}$. Therefore, item (i) is proved.

By rewriting (\ref{eq6}) as
   \begin{equation}\label{}
   V^{\alpha,\beta}(\rho,\Phi)=\frac{1+\sum_{i}\mathrm{tr}\rho^{\alpha+\beta}K_i \rho^ {1-\alpha-\beta} K_i^\dag}{2}-\sum^d_{l,m=1}\mathrm{tr}[\Phi(|l\rangle\langle m|\rho) |m\rangle\langle l| \rho], \notag
   \end{equation}
where $\alpha,\beta\geq 0,~\alpha+\beta\leq 1$ and
$\{|l\rangle\}^d_{l=1}$ is an orthonormal basis in $\mathcal{H}$,
it is not difficult to show that
$V^{\alpha,\beta}({\rho},\Phi)$ is positive-real-linear in $\Phi$.
Hence, item (ii) holds.

Let $X$ be a matrix, and $0\leq t\leq1$. By using
the Lieb's theorem \cite{NC,EHL}, the function
$f(A,B)=\mathrm{tr}(X^\dagger A^t XB^{1-t})$ is jointly concave in
positive matrices $A$ and $B$, which implies that
 $f\left(\sum_j\lambda_jA_j,\sum_j\lambda_jB_j\right)\geq\sum_j\lambda_jf(A_j,B_j)$, where $\lambda_j\geq 0$ with $\sum_j\lambda_j=1$, and $A_j$, $B_j$ be positive matrices for each $j$.
Taking $A_j=B_j=\rho_j, t=\alpha+\beta$, and $X=K_i$, we obtain
$\mathrm{tr}\left[(\sum_j\lambda_j\rho_j)^{\alpha+\beta}K_i(\sum_j\lambda_j\rho_j)^{1-\alpha-\beta}K_i^\dag\right]\geq\sum_j\lambda_j\mathrm{tr}\rho_j^{\alpha+\beta}K_i\rho_j^{1-\alpha-\beta}K_i^\dag$
for each $i$.  Note that
$|\mathrm{tr}(\sum_j\lambda_j\rho_j)K_i|^2\leq\sum_j\lambda_j|\mathrm{tr}\rho_jK_i|^2$
holds for each $i$. Summing over $i$ on both sides of the two
inequalities, item (iii) follows immediately.

Noting that $(U\rho U^\dag)^{\alpha+\beta}=U\rho^{\alpha+\beta}U^\dag$ and $(U\rho U^\dag)^{1-\alpha-\beta}=U\rho^{1-\alpha-\beta}U^\dag$ for $\alpha,\beta\geq0$ with $\alpha+\beta\leq1$, by Eq. (\ref{eq6}) and the cyclicity of the trace, item (iv) can be derived.

Suppose that $\Phi^A(\rho)=\sum_{i}K_i^A\rho
{K_i^A}^\dag$. Direct calculation shows that
\begin{align*}
&\mathrm{tr}(\rho^{A}\otimes\rho^{B})^{\alpha+\beta}(\Phi^A\otimes\mathcal{I}^B)((\rho^{A}\otimes\rho^{B})^{1-\alpha-\beta})
\\ \notag
=&
\mathrm{tr}({(\rho^A)}^{\alpha+\beta}\otimes{(\rho^B)}^{\alpha+\beta})
(\Phi^A({(\rho^A)}^{1-\alpha-\beta})\otimes{(\rho^B)}^{1-\alpha-\beta})
\notag \\
=&
\mathrm{tr}({(\rho^A)}^{\alpha+\beta}(\Phi^A({(\rho^A)}^{1-\alpha-\beta})\otimes\rho^B)
\notag \\
=&
\mathrm{tr}{(\rho^A)}^{\alpha+\beta}\Phi^A({(\rho^A)}^{1-\alpha-\beta}),
\end{align*}
and $\mathrm{tr}(\rho^A\otimes
\rho^B)(K_i^{A}\otimes\mathbf{1}^B)=\mathrm{tr}(\rho^AK_i^{A}\otimes\rho^B)=\mathrm{tr}(\rho^AK_i^{A})$.
Thus, it follows from Eq. (\ref{eq6}) that item(v) holds.

For item (vi), let $W$ be any operator on
$\mathcal{H}_{A}\otimes\mathcal{H}_{B}$, then
$\mathrm{tr}((F_A\otimes \mathbf{1}_B)W)=\mathrm{tr}(F_A\cdot
\mathrm{tr}_BW)$ for any $F_A$ on $\mathcal{H}_{A}$, where
$\mathbf{1_B}$ denotes the identity operator on
$\mathcal{H}_B$\cite{BHATIA}. Thus
\begin{align*}
&V^{\alpha,\beta}(\rho^{AB},\Phi^A\otimes\mathcal{I}^B)   \\ \notag
=&\frac{1+\sum_{i}\mathrm{tr}(\rho^{AB})^{\alpha+\beta}(K_i^A\otimes\mathbf{1}_B)(\rho^{AB})^{1-\alpha-\beta}
({K_i^A}^\dag\otimes\mathbf{1}_B)}{2}-\sum_{i}{|\mathrm{tr}\rho^{AB}(K_i^A\otimes\mathbf{1}_B)|}^{2}
\notag \\
\leq &
\frac{1+\sum_{i}\mathrm{tr}(\rho^A)^{\alpha+\beta}K_i^A(\rho^A)^{1-\alpha-\beta}{K_i^A}^\dag}
{2}-\sum_{i}{|\mathrm{tr}\rho^A {K_i^A}^\dag|}^{2}
\notag \\
=&V^{\alpha,\beta}(\rho^A,\Phi^A) ,~~~\alpha,\beta\geq
0,~\alpha+\beta\leq1,
\end{align*} where ${K_i^A}^\dag$ are the Kraus operators of the channel $\Phi^A$, and the last inequality follows from Lieb's concavity theorem\cite{EHL}.

Combining items (ii) and (vi), item (vii) follows
immediately.

The quantum uncertainty of $\Phi$ in $\rho$ is defined by
\begin{eqnarray}\label{eq8}
Q^{\alpha,\beta}({\rho},\Phi)=\sum_{i}Q^{\alpha,\beta}({\rho},K_i),
~~~\alpha,\beta\geq 0,~\alpha+\beta\leq 1.
\end{eqnarray}
By using Eq. (\ref{eq3}), $Q^{\alpha,\beta}({\rho},\Phi)$ can be further rewritten as,
\begin{align}\label{eq9}
Q^{\alpha,\beta}(\rho,\Phi)
=&\frac{1}{2}\sum_{i}(\mathrm{tr}\rho K_i^{\dag}K_i-\mathrm{tr}\rho^{\alpha}K_i\rho^{1-\alpha}K_i^\dag-\mathrm{tr}\rho^{\beta}K_i\rho^{1-\beta}K_i^\dag+\mathrm{tr}\rho^{\alpha+\beta}K_i\rho^{1-\alpha-\beta}K_i^\dag)  \notag \\
=&\frac{1}{2}[1-\mathrm{tr}\rho^{\alpha}\Phi(\rho^{1-\alpha})-\mathrm{tr}\rho^{\beta}\Phi(\rho^{1-\beta})+\mathrm{tr}\rho^{\alpha+\beta}\Phi(\rho^{1-\alpha-\beta})],
\end{align}where $\alpha,\beta\geq 0,~\alpha+\beta\leq 1$,
which could be viewed as a family of coherence measures with respect
to a channel $\Phi$ under certain restrictive conditions\cite{WZFL}.
When $\alpha=\beta=\frac{1}{2}$,
$Q^{\alpha,\beta}(\rho,\Phi)$ reduces to
$$Q(\rho,\Phi)=\sum_iQ(\rho,K_i)=\frac{1+\mathrm{tr}\rho\Phi(\mathbf{1})}{2}
-\mathrm{tr}\sqrt{\rho}\Phi(\sqrt{\rho}),$$
which has been introduced in
\cite{SL,SLYS} (up to a constant factor in \cite{SLYS}). Based on
the interpretation of $Q^{\alpha,\beta}({\rho},K)$,
$Q^{\alpha,\beta}({\rho},\Phi)$ can be interpreted as a family of
quantifiers of the information content of channels skew to $\Phi$ in
the state $\rho$.

The quantity $Q({\rho},\Phi)$ (denoted as
$I(\rho,\Phi)$ in \cite{SLYS}), which arises naturally from
algebraic and geometric manipulation of state-channel interaction,
has intrinsic informational-theoretical meaning. In fact, it can be
intuitively interpreted as the asymmetry, coherence, noncommutativity,
quantumness and quantum uncertainty (of state $\rho$ with respect to
channel $\Phi$) \cite{SLYS}. By replacing the commutator with
anti-commutator in $Q(\rho,\Phi)$, the quantity $J(\rho,\Phi)$, the dual to $Q(\rho,\Phi)$
($I(\rho,\Phi)$), has also been defined and interpreted as the symmetry,
incoherence, commutativity, classicality and classical uncertainty
(of state $\rho$ with respect to channel $\Phi$) \cite{SLYS}. The
symmetry-asymmetry complementarity relations have been derived and
applied to quantification of the degree of symmetry and
wave-particle duality\cite{SLYS,WZFL}. The generalized quantity
$Q^{\alpha,\beta}({\rho},\Phi)$ can therefore serve as a family of
quantifiers of the measures, similar to the interpretations of
$Q({\rho},\Phi)$. Note that in this paper, we follow the lines of
\cite{SL} and identify $Q^{\alpha,\beta}({\rho},\Phi)$ with quantum
uncertainty, whereas the classical uncertainty
$C^{\alpha,\beta}({\rho},\Phi)$ is identified by fixing the
difference of the total uncertainty (quantified by MGV
$V^{\alpha,\beta}({\rho},\Phi)$) and the quantum uncertainty, which
is a little different from the formulations in \cite{SLYS}.

{\bf Remark 2.1} Note that $Q(\rho,\Phi)$ is
ancillary independent, decreasing under partial trace and
superadditive\cite{SLYS}. The ancillary independence (in a strong
version, which is in fact invariance under partial trace) and
additivity of $V(\rho,\Phi)$ has been proved in \cite{SL}. However,
for $V^{\alpha,\beta}(\rho,\Phi)$ with $\alpha,\beta\geq0$ and
$\alpha+\beta\leq1$, we can only prove that it is ancillary
independent (in the sense of \cite{SLYS}), increasing under partial
trace and subadditive.

We define the classical uncertainty of $\Phi$ in $\rho$ as
\begin{equation}\label{eq10}
C^{\alpha,\beta}({\rho},\Phi) =\sum_{i}C^{\alpha,\beta}({\rho},K_i),
~~~\alpha,\beta\geq 0,~\alpha+\beta\leq 1.
\end{equation}
By using Eq. (\ref{eq4}), $C^{\alpha,\beta}({\rho},\Phi)$ can be further rewritten as,
\begin{align}\label{eq11}
C^{\alpha,\beta}({\rho},\Phi) \notag
=&\frac{1}{2}\sum_{i}(\mathrm{tr}\rho^ {\alpha}K_i\rho^{1-\alpha}K_i^\dag+\mathrm{tr}\rho^ {\beta}K_i\rho^{1-\beta}K_i^\dag)-\sum_{i}{|\mathrm{tr}\rho K_i|}^{2} \\
=&\frac{1}{2}[\mathrm{tr}\rho^{\alpha}\Phi(\rho^{1-\alpha})+\mathrm{tr}\rho^{\beta}\Phi(\rho^{1-\beta})]-\sum_{i}{|\mathrm{tr}\rho K_i|}^{2},
\end{align}where $\alpha,\beta\geq 0,~\alpha+\beta\leq 1$.
 When $\alpha=\beta=\frac{1}{2}$,
$C^{\alpha,\beta}(\rho,\Phi)$ reduces to $C(\rho,\Phi)$, and Eq. (\ref{eq11}) reduces to the following equation in \cite{SL}, i.e.,
\begin{align}\label{}
C({\rho},\Phi)=\sum_{i}C(\rho,K_i)=\mathrm{tr}\sqrt{\rho}\Phi(\sqrt{\rho})-\sum_{i}{|\mathrm{tr}\rho K_i|}^{2}. \notag
\end{align}
It can also be verified that the quantities $Q^{\alpha,\beta}({\rho},\Phi)$ and $C^{\alpha,\beta}({\rho},\Phi)$ defined in Eqs. (\ref{eq8}) and (\ref{eq10}) are independent of the choice of the Kraus operators of $\Phi$.
 Consequently we have
 \begin{equation}\label{eq12}
  V^{\alpha,\beta}({\rho},\Phi)=Q^{\alpha,\beta}({\rho},\Phi)+C^{\alpha,\beta}({\rho},\Phi),~~~\alpha,\beta\geq 0,~\alpha+\beta\leq 1,
  \end{equation}
which means that the total uncertainty of a quantum channel can also
be decomposed into the quantum and classical counterparts.
As is shown in \cite{SL}, we can similarly verify
that $C^{\alpha,\beta}({\rho},\Phi)=0$ when
$\rho=|\psi\rangle\langle \psi|$ is a pure state, and in this case
 \begin{align} \label{eq13}
 V^{\alpha,\beta}({\rho},\Phi)=Q^{\alpha,\beta}({\rho},\Phi)
 =&\frac{1}{2}\left(1-\sum_i\langle
\psi| K_i|\psi\rangle \langle
\psi| K_i^{\dag}|\psi\rangle \right)  \notag \\
=& \frac{1}{2}\left(1-\sum_i \langle K_i \rangle \langle K_i^{\dag} \rangle\right)\notag \\
=& \frac{1}{2}\left(1-\sum_i |\langle K_i \rangle|^2 \right),
\end{align}
where $\langle K_i\rangle$ denotes the expectation value of $K_i$
for each $i$. Suppose that $\rho$ is a state on quantum system $B$
and $A$ is an ancillary system. Define a pure state
${|\psi^{AB}}\rangle$ for the joint system $AB$ such that
$\mathrm{tr}_A|\psi^{AB}\rangle\langle \psi^{AB}|=\rho$.
By Eq. (\ref{eq13}) and property (vi) of
$V^{\alpha,\beta}({\rho},\Phi)$, we have
 \begin{equation}\label{I}
 V^{\alpha,\beta}({\rho},\Phi)\geq V^{\alpha,\beta}(|\psi^{AB}\rangle\langle \psi^{AB}|,\mathcal{I}^A\otimes\Phi)=Q^{\alpha,\beta}(|\psi^{AB}\rangle\langle \psi^{AB}|,\mathcal{I}^A\otimes\Phi). \notag
 \end{equation}
It shows that the total uncertainty of a quantum channel $\Phi$ in a
state $\rho$ is no less than the quantum uncertainty of this channel
in the corresponding purified states ${|\psi^{AB}}\rangle$.

\vskip0.1in

\noindent {\bf 3 A trade-off relation between MGV-based total uncertainty of quantum channels and entanglement fidelity}\\\hspace*{\fill}\\
In this section, we first recall the concepts of entanglement
fidelity, entropy exchange, and coherent information. We then derive
a trade-off relation between the MGV-based total uncertainty of
quantum channels and the entanglement fidelity. Furthermore, we exploit
this trade-off relation to explore the relations between the
MGV-based total uncertainty of quantum channels and entropy
exchange/coherent information.

For a quantum state $\rho$ and a quantum channel $\Phi$ on system
$B$, the entanglement fidelity is defined as \cite{NC,SB1}
\begin{align}\label{eq14}
F_e({\rho},\Phi)
=&F(|\psi^{AB}\rangle,\mathcal{I}^A\otimes\Phi(|\psi^{AB}\rangle\langle \psi^{AB}|))^{2}   \notag \\
=&\langle
\psi^{AB}|\mathcal{I}^A\otimes\Phi(|\psi^{AB}\rangle\langle
\psi^{AB}|)|\psi^{AB}\rangle,
\end{align}
where $A$ is an auxiliary system, $|\psi^{AB}\rangle$ is a
purification of $\rho$ satisfying
$\mathrm{tr}_A|\psi^{AB}\rangle\langle \psi^{AB}|=\rho$, and
$F(\sigma_1, \sigma_2)=\mathrm{tr}\sqrt{\sqrt{\sigma_1} \sigma_2
\sqrt{\sigma_1}}$ is the quantum fidelity between quantum states
$\sigma_1$ and $\sigma_2$. The entanglement fidelity provides a
quantification of to what extent the entanglement of
$|\psi^{AB}\rangle$ can be preserved under the quantum channel $\Phi$.
$F_e({\rho},\Phi)$ does not depend on the ways of purification and can be rewritten as \cite{NC},
\begin{equation}\label{eq15}
F_e({\rho},\Phi)=\sum_{i}{|\mathrm{tr}\rho K_i|}^{2}.
\end{equation}

For a quantum state $\rho$ and a quantum channel $\Phi$ on system
$B$, the entropy exchange is defined as \cite{SB1}
\begin{equation}\label{eq16}
S_e({\rho},\Phi)=S({\rho^{A'B'}})=-\mathrm{tr}{\rho^{A'B'}}\mathrm{log}{\rho^{A'B'}},
\end{equation}
where
${\rho^{A'B'}}=\mathcal{I}^A\otimes\Phi{(|\psi^{AB}\rangle\langle
\psi^{AB}|)}$, $|\psi^{AB}\rangle$ is a purification of $\rho$ with
auxiliary system $A$, $S(\sigma)=-\mathrm{tr}{\sigma}\mathrm{log}{\sigma}$ is the von
Neumann entropy of a quantum state $\sigma$, and the logarithm `log' is taken to be base 2.
The entropy exchange quantifies the amount of information exchanged
between system $B$ and the environment under the action of channel
$\Phi$. Correspondingly, the coherent information is defined as \cite{SB2}
\begin{equation}\label{eq17}
I_c({\rho},\Phi)=S(\rho')-S_e=S(\rho')-S(\rho^{A'B'}),
\end{equation}
where $\rho'=\Phi(\rho)$ is the state of system $B$ under the action
of channel $\Phi.$  $I_c({\rho},\Phi)$ identifies how much
quantum information is transmitted when the quantum channel is applied.

Utilizing Eqs. (\ref{eq6}) and (\ref{eq15}), we obtain the following trade-off relation between the MGV-based total uncertainty of quantum channels and the entanglement fidelity,
\begin{align} \label{eq18}
 V^{\alpha,\beta}({\rho},\Phi)+F_e({\rho},\Phi)
 =&\frac{1+\mathrm{tr}\rho^{\alpha+\beta}\Phi (\rho^ {1-\alpha-\beta})}{2}  \notag \\
\leq& \frac{1+\lambda_{\mathrm{max}}(\Phi (\rho^ {1-\alpha-\beta}))
\mathrm{tr}\rho^{\alpha+\beta}}{2},
\end{align}
where $\alpha,\beta\geq 0,~\alpha+\beta\leq 1$, and
$\lambda_{\mathrm{max}}(\cdot)$ denotes the maximum spectrum of a
matrix. This trade-off relation is an extension of Eq. (\ref{eq4})
in \cite{SL}, showing that the entanglement
fidelity $F_e({\rho},\Phi)$ cannot be too large if the total
uncertainty of $\Phi$ in $\rho$ is large. We find that if the
nonzero eigenvalues of $\Phi(\rho^ {1-\alpha-\beta})$ are all equal,
then the inequality (\ref{eq18}) is saturated. Note that if
$\alpha+\beta=1$ and $\Phi$ is unital, we get
$$V(\rho,\Phi)+F_e(\rho,\Phi)=1,$$
which has been established in Ref.\cite{SL}, and may be viewed as a
demonstration of certain information conservation.

For pure states $\rho=|\psi\rangle\langle\psi|$, we
have $\mathrm{tr}\rho^{\alpha+\beta}\Phi (\rho^
{1-\alpha-\beta})=\sum_{i}{|\langle\psi|K_i|\psi\rangle|}^{2}=\sum_{i}{|\mathrm{tr}\rho
K_i|}^{2}\\=F_e(\rho,\Phi)$. Thus Eq. (\ref{eq18}) can be
rewritten as the following trade-off relation,
\begin{equation*}
2V^{\alpha,\beta}(|\psi\rangle\langle\psi|,\Phi)+F_e(|\psi\rangle\langle\psi|,\Phi)=1.
\end{equation*}
It can be seen that the total uncertainty
$V^{\alpha,\beta}(|\psi\rangle\langle\psi|,\Phi)$ (or equivalently,
the quantum uncertainty
$Q^{\alpha,\beta}(|\psi\rangle\langle\psi|,\Phi)$, which is due to
Eq. (\ref{eq13})) characterizes the information loss associated with
the entanglement of the initial purified state $|\psi^{AB}\rangle$.

We now apply these trade-off relations to deduce
the connections between the total uncertainty of quantum channels
and the entropy exchange/coherent information as a consequence of
the above trade-off relation. Recall that the
quantum Fano inequality is $S_e(\rho,\Phi)\leq
H(F_e(\rho,\Phi))+(1-F_e(\rho,\Phi))\mathrm{log}(d^2-1)$, where
$H(\cdot)$ is the binary Shannon entropy, i.e.,
$H(p)=-p\mathrm{log}p-(1-p)\mathrm{log}(1-p)$ for $0\leq p \leq
1$\cite{NC}. Combining this inequality, Eq. (\ref{eq18}), and the
facts that $H(F_e(\rho,\Phi))\leq1$ and
$\mathrm{log}(d^2-1)\leq 2\mathrm{log}d$, we have
\begin{eqnarray*}
S_e({\rho},\Phi)
&\leq& H\left(\frac{1+\mathrm{tr}\rho^{\alpha+\beta}\Phi (\rho^{1-\alpha-\beta})-2V^{\alpha,\beta}({\rho},\Phi)}{2}\right)
\\&+&\left(\frac{2V^{\alpha,\beta}({\rho},\Phi)+1-\mathrm{tr}\rho^{\alpha+\beta}\Phi (\rho^ {1-\alpha-\beta})}{2}\right)
\mathrm{log}(d^2-1)
\\&\leq& 1+(2V^{\alpha,\beta}({\rho},\Phi)+1-\mathrm{tr}\rho^{\alpha+\beta}\Phi (\rho^ {1-\alpha-\beta}))\mathrm{log}d,
\end{eqnarray*}
where $\alpha,\beta\geq 0$ and $\alpha+\beta\leq 1$.
Therefore, the MGV-based total uncertainty of quantum channels
$V^{\alpha,\beta}({\rho},\Phi)$ can serve as an upper bound of the
entropy exchange $S_e({\rho},\Phi)$,
\begin{equation}\label{eq19}
1+(2V^{\alpha,\beta}({\rho},\Phi)+1-\mathrm{tr}\rho^{\alpha+\beta}\Phi (\rho^ {1-\alpha-\beta}))\mathrm{log}d\geq S_e({\rho},\Phi),
\end{equation}
where $\alpha,\beta\geq 0,~\alpha+\beta\leq 1$, and
$d$ is the dimension of the system Hilbert space. This demonstrates
that if the information exchange with environment is large, the
total uncertainty of quantum channels cannot be arbitrarily small.
Note that when $\alpha+\beta=1$ and $\Phi$ is unital, we get
$$1+2V({\rho},\Phi)\mathrm{log}d\geq S_e({\rho},\Phi),$$
which has been established in Ref.\cite{SL}.

The quantum Fano inequality reveals that if the
entropy exchange for a process is large, then the entanglement
fidelity for the process must necessarily be small, indicating that
the entanglement between $A$ and $B$ has not been well
preserved\cite{NC}. This inequality is saturated if
$\rho^{A'B'}=\mathrm{diag}(p_1,p_2,\cdots,p_{d^2})$, where
$p_1=F_e(\rho,\Phi)$ and $p_i=(1-p_1)/(d^2-1)$ for
$i=2,\cdots,d^2$\cite{NC}. Note that $H(F_e(\rho,\Phi))$ attains its
maximum value of $1$ at $F_e(\rho,\Phi)=1/2$, while
$\mathrm{log}(d^2-1)\leq 2\mathrm{log}d$ cannot be saturated for any
finite number $d$, but the values of $\mathrm{log}(d^2-1)$ and
$2\mathrm{log}d$ can be arbitrarily close when $d$ is sufficiently
large. This implies that if
$\rho^{A'B'}=\mathrm{diag}(1/2,1/2(d^2-1),\cdots,1/2(d^2-1))$ and
$d$ is very large, we have
$$S_e({\rho},\Phi)=1+\frac{1}{2}\mathrm{log}(d^2-1)\approx
1+\mathrm{log}d,$$ and in this case,
$V^{\alpha,\beta}({\rho},\Phi)=\frac{1}{2}\mathrm{tr}\rho^{\alpha+\beta}\Phi
(\rho^ {1-\alpha-\beta})$. Thus, we can say that Eq.
(\ref{eq19}) is saturated approximately in a high-dimensional
Hilbert space (the dimension $d$ is large enough) when
$\rho^{A'B'}=\mathrm{diag}(1/2,1/2(d^2-1),\cdots,1/2(d^2-1))$.

From Eqs. (\ref{eq17}) and (\ref{eq19}), we obtain
\begin{equation}
1+(2V^{\alpha,\beta}({\rho},\Phi)+1-\mathrm{tr}\rho^{\alpha+\beta}\Phi (\rho^ {1-\alpha-\beta}))\mathrm{log}d+I_c(\rho,\Phi)\geq S(\rho'), \notag
\end{equation}
where $\alpha,\beta\geq 0,~\alpha+\beta\leq 1$.

Combining Eq. (\ref{eq19}) and the above inequality,
we obtain
\begin{align*}
&2+2(2V^{\alpha,\beta}({\rho},\Phi)+1-\mathrm{tr}\rho^{\alpha+\beta}\Phi (\rho^ {1-\alpha-\beta}))\mathrm{log}d+I_c(\rho,\Phi)\notag \\
\geq& S_e(\rho,\Phi)+S(\rho')  \notag \\
\geq& S(\rho)+2S_e(\rho,\Phi)  \notag \\
\geq& S(\rho), \notag
\end{align*}
where the second inequality follows from the fact that
$S(\rho')-S(\rho)\geq S_e(\rho,\Phi)$ \cite{NC,SB2}, and we can
further obtain
\begin{equation}\label{eq20}
2(2V^{\alpha,\beta}({\rho},\Phi)+1-\mathrm{tr}\rho^{\alpha+\beta}\Phi
(\rho^ {1-\alpha-\beta}))\mathrm{log}d+I_c(\rho,\Phi)\geq S(\rho)-2,
\end{equation}
where $\alpha,\beta\geq 0,~\alpha+\beta\leq 1$, which gives a
relation between the MGV-based total uncertainty of quantum channels
and the coherent information. This is a
manifestation that the total uncertainty of quantum channels cannot
be arbitrarily small when the coherent information is very small.
Note that when $\alpha+\beta=1$ and $\Phi$ is unital, we get
$$4V({\rho},\Phi)\mathrm{log}d+I_c(\rho,\Phi)\geq S(\rho)-2,$$
which has been established in Ref.\cite{SL}.

From the proof of Eq. (20), it can be easily seen
that it is saturated approximately in a high-dimensional Hilbert
space (the dimension $d$ is large enough) when
$\rho^{A'B'}=\mathrm{diag}(1/2,1/2(d^2-1),\cdots,1/2(d^2-1))$ and
$\Phi$ preserves the von Neumann entropy (i.e.,
$S(\rho')=S(\rho)$).

\vskip0.1in

\noindent {\bf 4 Examples and an experimental protocol}\\\hspace*{\fill}\\
In this section, we calculate the total uncertainty
$V^{\alpha,\beta}(\rho,\Phi)$ and the quantum uncertainty
$Q^{\alpha,\beta}(\rho,\Phi)$ for several typical quantum channels,
and illustrate the trade-off relation
(\ref{eq18}), inequalities (\ref{eq19}) and (\ref{eq20}) for specific channels
and states.

A qubit state can be written as
$\rho=\frac{1}{2}(\mathbf{1}+\mathbf{r}\cdot\bm{\sigma})$, where
$\mathbf{r}=(r_1,r_2,r_3)$ is the Bloch vector satisfying
$|\mathbf{r}|\leq 1$, $\bm{\sigma}=(\sigma_1,\sigma_2,\sigma_3)$
with $\sigma_j$ $(j=1,2,3)$ the Pauli matrices, and
$\mathbf{r}\cdot\bm{\sigma}=\sum^3_{j=1}r_j\sigma_j$. Denote
$r=|\mathbf{r}|$. Then the eigenvalues of $\rho$ are
$\lambda_{1,2}=(1\mp r)/2$ and \cite{XKW}
$$\rho^{\kappa}=\left(\begin{array}{cc}
\frac{\lambda_1^{\kappa}+\lambda_2^{\kappa}}{2}+\frac{r_3(\lambda_2^{\kappa}
-\lambda_1^{\kappa})}{2r}&\frac{(-r_1+ir_2)(\lambda_1^{\kappa}-\lambda_2^{\kappa})}{2r}\\
\frac{(-r_1-ir_2)(\lambda_1^{\kappa}-\lambda_2^{\kappa})}
{2r}&\frac{\lambda_1^{\kappa}+\lambda_2^{\kappa}}{2}
-\frac{r_3(\lambda_2^{\kappa}-\lambda_1^{\kappa})}{2r}\\
\end{array}
\right).
$$

\vskip0.1in
\noindent {\bf Example 1} Consider the amplitude damping channel $\Phi(\rho)=\sum_{i=1}^2K_i\rho K_i^\dag$ with the
Kraus operators
$$K_1=\left(\begin{array}{cc}
1&0\\
0&\sqrt{1-p}\\
\end{array}
\right),~~
K_2=\left(\begin{array}{cc}
0&\sqrt{p}\\
0&0\\
\end{array}
\right),~~
0\leq p\leq 1.
$$
The total uncertainty and quantum uncertainty of $\Phi$ for the
qubit state $\rho$ are
\begin{eqnarray}\label{eq21}
&&V^{\alpha,\beta}(\rho,\Phi)\nonumber\\
&=&\left[\frac{1}{4}(\lambda_1^{\alpha+\beta}+\lambda_2^{\alpha+\beta})(\lambda_1^{1-\alpha-\beta}+\lambda_2^{1-\alpha-\beta})-\frac{pr^2}{4}+\frac{(p+\sqrt{1-p}-1)}{2}{r_3^2}\right.
\nonumber\\
&&\left.-\frac{2pr_3-p+2\sqrt{1-p}}{4}+\frac{pr_3}{4r}(\lambda_1^{1-{\alpha-\beta}}\lambda_2^{\alpha+\beta}-\lambda_2^{1-{\alpha-\beta}}\lambda_1^{\alpha+\beta}+\lambda_2-\lambda_1)\right.
\nonumber\\
&&\left.+\frac{[2-p-pr_3^2+2\sqrt{1-p}(r^2-r_3^2)]}{8r^2}(\lambda_1^{\alpha+\beta}
-\lambda_2^{\alpha+\beta})(\lambda_1^{1-{\alpha-\beta}}
-\lambda_2^{1-{\alpha-\beta}})\right]
\end{eqnarray}
and
\begin{eqnarray}\label{eq22}
Q^{\alpha,\beta}(\rho,\Phi)
&=&\frac{1}{2}\left[\frac{(1-\sqrt{1-p})(r_1^2+r_2^2)+pr_3^2}{2r^2}(\lambda_1^{1-\alpha-\beta}+\lambda_2^{1-\alpha-\beta})\right.
\nonumber\\
&&\left.+\frac{pr_3}{2r}(\lambda_1^{1-\alpha-\beta}-\lambda_2^{1-\alpha-\beta})\right]
(\lambda_1^{\alpha}-\lambda_2^{\alpha})(\lambda_1^{\beta}-\lambda_2^{\beta}),
\end{eqnarray}
respectively, where $\alpha,\beta\geq 0$ and $\alpha+\beta\leq 1$.

\vskip0.1in

\noindent {\bf Example 2} Consider the phase damping channel $\Phi(\rho)=\sum_{i=1}^2K_i\rho K_i^\dag$ with the
Kraus operators
$$K_1=\left(\begin{array}{cc}
1&0\\
0&\sqrt{1-p}\\
\end{array}
\right),~~
K_2=\left(\begin{array}{cc}
0&0\\
0&\sqrt{p}\\
\end{array}
\right),~~
0\leq p\leq 1.
$$
The total uncertainty and quantum uncertainty of $\Phi$ for the
qubit state $\rho$ are
\begin{eqnarray}\label{eq23}
&&V^{\alpha,\beta}(\rho,\Phi)\nonumber\\
&=&\left[\frac{1}{4}(\lambda_1^{\alpha+\beta}
+\lambda_2^{\alpha+\beta})(\lambda_1^{1-\alpha-\beta}
+\lambda_2^{1-\alpha-\beta})-\frac{[\sqrt{1-p}+(1-\sqrt{1-p})r_3^2]}{2}
\right.
\nonumber\\
&&\left.+\frac{[(1-\sqrt{1-p})r_3^2+\sqrt{1-p}r^2]}{4r^2}(\lambda_1^{\alpha+\beta}
-\lambda_2^{\alpha+\beta})(\lambda_1^{1-\alpha-\beta}-\lambda_2^{1-\alpha-\beta})
\right]
\end{eqnarray}
and
\begin{eqnarray}\label{eq24}
&&Q^{\alpha,\beta}(\rho,\Phi)\nonumber \\
&=&\frac{1}{2}\frac{(1-\sqrt{1-p})(r_1^2+r_2^2)}{2r^2}
(\lambda_1^{\alpha}-\lambda_2^{\alpha})(\lambda_1^{\beta}-\lambda_2^{\beta})(\lambda_1^{1-\alpha-\beta}+\lambda_2^{1-\alpha-\beta}),
\end{eqnarray}
respectively, where $\alpha,\beta\geq 0$ and $\alpha+\beta\leq 1$.

\vskip0.1in

\noindent {\bf Example 3} For the depolarizing channel
\begin{equation}
\Phi(\rho)=(1-3p)\rho+p\sum_{j=1}^3 \sigma_j\rho\sigma_j,~\,\,0\leq p\leq \frac{1}{3}\notag,
\end{equation}
the total uncertainty and quantum uncertainty of $\Phi$ for the qubit state
$\rho$ are given by
\begin{eqnarray}\label{eq25}
V^{\alpha,\beta}(\rho,\Phi)
&=&\frac{1}{4}\left[2+(\lambda_1^{\alpha+\beta}+\lambda_2^{\alpha+\beta})(\lambda_1^{1-\alpha-\beta}+\lambda_2^{1-\alpha-\beta})\right.
\nonumber\\
&&\left.+(1-4p)(\lambda_1^{\alpha+\beta}-\lambda_2^{\alpha+\beta})(\lambda_1^{1-\alpha-\beta}-\lambda_2^{1-\alpha-\beta})\right]-(1-3p+pr^2)
\end{eqnarray}
and
\begin{equation}\label{eq26}
Q^{\alpha,\beta}(\rho,\Phi) =p
(\lambda_1^{\alpha}-\lambda_2^{\alpha})(\lambda_1^{\beta}-\lambda_2^{\beta})(\lambda_1^{1-\alpha-\beta}+\lambda_2^{1-\alpha-\beta}),
\end{equation}
respectively, where $\alpha,\beta\geq 0$ and $\alpha+\beta\leq 1$.

\vskip0.1in

\noindent {\bf Example 4} The  completely decoherent channel \cite{BCD,BHT} is given by $\Phi(\rho)=M\circ\rho$, where $M$ is a correlation matrix
which is positive semidefinite with all the diagonal elements $1$, and $\circ$ denotes the Hardamard product of matrices. For the following $2\times 2$ correlation matrix,
$$
M=\left(\begin{array}{cc}
1&\theta\\
\theta&1\\
\end{array}
\right),~~
-1\leq\theta\leq 1,
$$
we have the total uncertainty and quantum uncertainty of $\Phi$ for the
qubit state $\rho$,
\begin{eqnarray}\label{eq27}
V^{\alpha,\beta}(\rho,\Phi)
&=& \left[\frac{1}{4}(\lambda_1^{\alpha+\beta}+\lambda_2^{\alpha+\beta})(\lambda_1^{1-\alpha-\beta}+\lambda_2^{1-\alpha-\beta})-\frac{{r_3}^2+ \theta-{r_3}^2\theta}{2} \right.
\nonumber\\
&&\left.+\frac{1}{r^2}(\lambda_1^{\alpha+\beta}-\lambda_2^{\alpha+\beta})(\lambda_1^{1-\alpha-\beta}
-{\lambda_2^{1-\alpha-\beta})(\theta r^2+r_3^2-\theta r_3^2)}\right]
\end{eqnarray}
and
\begin{eqnarray}\label{eq28}
Q^{\alpha,\beta}(\rho,\Phi)
&=&\frac{(\lambda_1^{\alpha+\beta}+\lambda_2^{\alpha+\beta})(\lambda_1^{1-\alpha-\beta}
+\lambda_2^{1-\alpha-\beta})+2}{4}\nonumber\\
&&-\frac{(\lambda_1^{\alpha}+\lambda_2^{\alpha})(\lambda_1^{1-\alpha}
+\lambda_2^{1-\alpha})+(\lambda_1^{\beta}+\lambda_2^{\beta})(\lambda_1^{1-\beta}+\lambda_2^{1-\beta})}{4}\nonumber\\
&&-\frac{r_3^2+\theta(r_1^2+r_2^2)}{4r^2}[(\lambda_1^{\alpha}-\lambda_2^{\alpha})(\lambda_1^{1-\alpha}
-\lambda_2^{1-\alpha})
+(\lambda_1^{\beta}-\lambda_2^{\beta})(\lambda_1^{1-\beta}
-\lambda_2^{1-\beta})]\nonumber\\
&&+\frac{r_3^2+\theta(r_1^2+r_2^2)}{4r^2}(\lambda_1^{\alpha+\beta}-\lambda_2^{\alpha+\beta})(\lambda_1^{1-\alpha-\beta}
-\lambda_2^{1-\alpha-\beta}),
\end{eqnarray}
respectively, where $\alpha,\beta\geq 0$ and $\alpha+\beta\leq 1$.

\vskip0.1in

In the next example we plot the total uncertainty
$V^{\alpha,\beta}(\rho,\Phi)$ and quantum uncertainty
$Q^{\alpha,\beta}(\rho,\Phi)$ for two kinds of
important quantum states, the Werner and isotropic states.

\noindent {{\bf Example 5} Fix an orthonormal basis
$\{|l\rangle\}_{l=1}^{d}$ of a $d$-dimensional Hilbert space
$\mathcal{H}$, $\{|l\rangle\langle m|\}_{l,m=1}^{d}$ is an
orthonormal basis of $\mathcal{B(H)}$, with the Hilbert-Schmidt
inner product ${\langle A^{\dag}|B\rangle}$:=$\mathrm{tr}AB$. For
simplicity, denote by $\{X_i\}_{i=1}^{d^2}$ the orthonormal basis of
$\mathcal{B(H)}$. It can be verified that
$\Phi(\rho)=\sum_{i=1}^{d^2}X_i\rho X_i^\dag/d$ is a quantum
channel. For this quantum channel we have
\begin{equation}\label{eq29}
V^{\alpha,\beta}(\rho,\Phi)=\frac{1}{2d}[d+\mathrm{tr}{\rho}^{\alpha
+\beta}\mathrm{tr}{\rho}^{1-\alpha-\beta}-2\mathrm{tr}{\rho}^{2}]
\end{equation}
and
\begin{equation}\label{eq30}
Q^{\alpha,\beta}(\rho,\Phi)=\frac{1}{2d}[d+\mathrm{tr}{\rho}^{1-\alpha-\beta}
\mathrm{tr}{\rho}^{\alpha+\beta}-\mathrm{tr}{\rho}^{1-\alpha}\mathrm{tr}{\rho}^{\alpha}
-\mathrm{tr}{\rho}^{1-\beta}\mathrm{tr}{\rho}^{\beta}],
\end{equation}
where $\alpha,\beta\geq 0$ and $\alpha+\beta\leq 1$.

Now we derive the total uncertainty and quantum uncertainty in Eqs.
(\ref{eq29}) and (\ref{eq30}) for two classes of states, the Werner
state and the isotropic state, respectively. First consider the Werner state,
$$
\rho_w=\left(\begin{array}{cccc}
         \frac{1}{3}p&0&0&0\\
         0&\frac{1}{6}(3-2p)&\frac{1}{6}(4p-3)&0\\
         0&\frac{1}{6}(4p-3)&\frac{1}{6}(3-2p)&0\\
         0&0&0&\frac{1}{3}p\\
         \end{array}
         \right),
$$
where $p\in[0,1]$. Note that $\rho_w$ is separable when
$p\in[0,\frac{1}{3}]$. According to Eqs. (\ref{eq29}) and (\ref{eq30}), we obtain
\begin{eqnarray}\label{eq31}
&&V^{\alpha,\beta}(\rho_w,\Phi)\nonumber\\
&=&\frac{1}{8}\left[3+6p-\frac{8p^2}{3}+3^{\alpha+\beta}p^{1-\alpha-\beta}(1-p)^{\alpha+\beta}+3^{1-\alpha-\beta}p^{\alpha+\beta}(1-p)^{1-\alpha-\beta}\right]
\end{eqnarray}
and
\begin{eqnarray}\label{eq32}
&&Q^{\alpha,\beta}(\rho_w,\Phi)\nonumber\\
&=&\frac{1}{8}\left[3-2p-3^{1-\alpha}p^{\alpha}(1-p)^{1-\alpha}-3^{\alpha}p^{1-\alpha}(1-p)^{\alpha}-3^{1-\beta}p^{\beta}(1-p)^{1-\beta}\right.
\nonumber\\
&&\left.-3^{\beta}p^{1-\beta}(1-p)^{\beta}+3^{1-\alpha-\beta}p^{\alpha+\beta}(1-p)^{1-\alpha-\beta}+3^{\alpha+\beta}p^{1-\alpha-\beta}(1-p)^{\alpha+\beta}\right].
\end{eqnarray}
Fig.~\ref{fig:Fig1} illustrates the values of
$V^{\alpha,\beta}(\rho_w,\Phi)$ and $Q^{\alpha,\beta}(\rho_w,\Phi)$
given in Eqs. (\ref{eq31}) and (\ref{eq32}) with
$\alpha=\frac{1}{5}$ and $\beta=\frac{3}{10}$.
Direct calculation shows that
$V^{\alpha,\beta}(\rho_w,\Phi)-Q^{\alpha,\beta}(\rho_w,\Phi)$, i.e.,
the classical uncertainty $C^{\alpha,\beta}(\rho_w,\Phi)$ reaches
its maximum value of $0.9375$ when $p=\frac{3}{4}$,
$\alpha=0.120797$ and $\beta=0.650832$. Interestingly, the linear
entropy (mixedness) $1-\mathrm{tr}{\rho_w^2}=2p-\frac{4}{3}p^2$ of
the Werner state attains its maximum value of $\frac{3}{4}$ when
$p=\frac{3}{4}$. In Fig.~\ref{fig:Fig2}, we plot the surfaces of
total uncertainty $V^{\alpha,\beta}(\rho_w,\Phi)$, quantum
uncertainty $Q^{\alpha,\beta}(\rho_w,\Phi)$ in Eqs. (\ref{eq31}) and
(\ref{eq32}), and their gap, the classical uncertainty
$C^{\alpha,\beta}(\rho_w,\Phi)$, for fixed values of $p$.
\begin{figure}[ht]
\centering
{\begin{minipage}[XuCong-Uncertainty-MGV-1]{0.5\linewidth}
\includegraphics[width=0.95\textwidth]{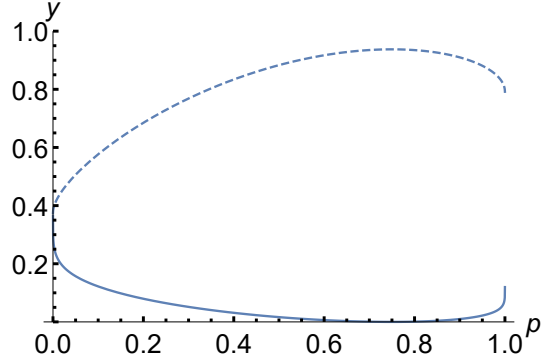}
\end{minipage}}
\caption{{The $y$-axis is the values of $V^{\alpha,\beta}(\rho_w,\Phi)$ and $Q^{\alpha,\beta} (\rho_w,\Phi)$.
 Dashed (solid) line represents the value of $V^{\alpha,\beta}(\rho_w,\Phi)$ ($Q^{\alpha,\beta}(\rho_w,\Phi)$) in Eq. (\ref{eq31}) (Eq. (\ref{eq32})) with  $\alpha=\frac{1}{5}$ and $\beta=\frac{3}{10}$. Note that the Werner state $\rho_w$ degenerates to a pure state when $p=0$ and in this case $V^{\alpha,\beta}(\rho_w,\Phi)=Q^{\alpha,\beta}(\rho_w,\Phi)$.\label{fig:Fig1}}}
\end{figure}

\begin{figure}[H]\centering
\subfigure[]
{\begin{minipage}[2-a-XuCong-Uncertainty-MGV]{0.24\linewidth}
\includegraphics[width=1.0\textwidth]{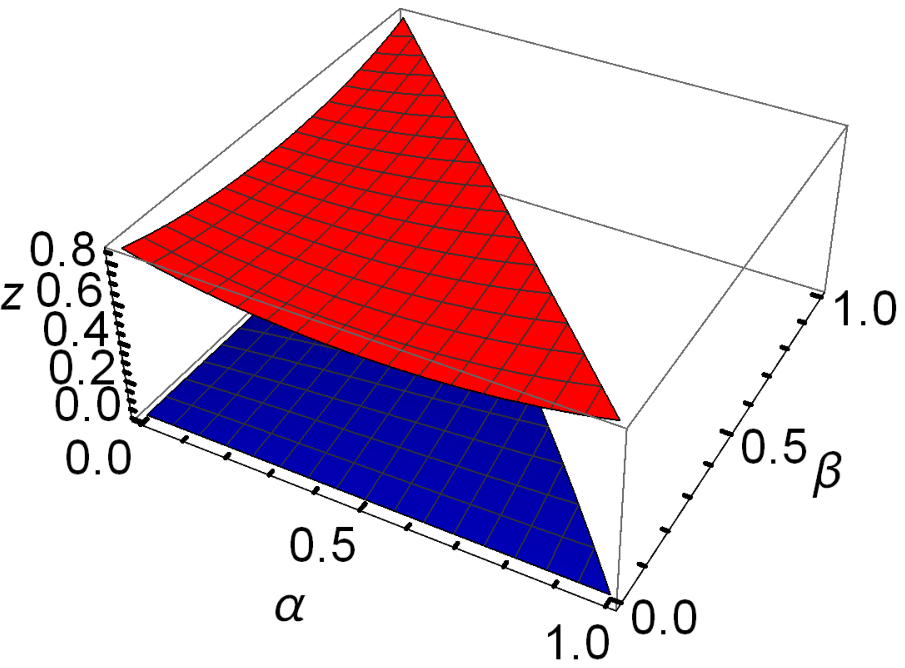}
\end{minipage}}
\subfigure[]
{\begin{minipage}[2-b-XuCong-Uncertainty-MGV]{0.24\linewidth}
\includegraphics[width=1.0\textwidth]{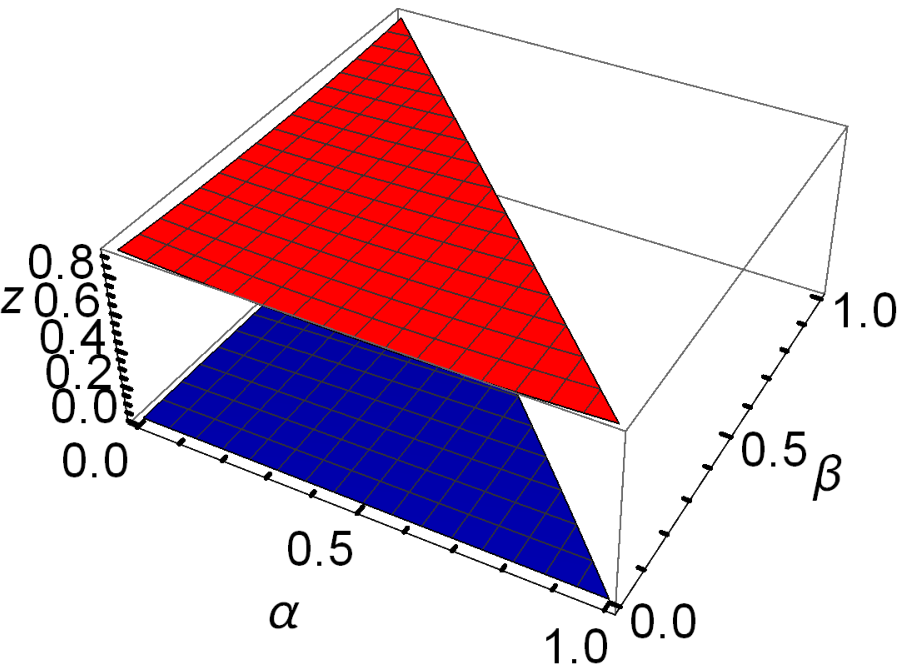}
\end{minipage}}
\subfigure[]
{\begin{minipage}[2-c-XuCong-Uncertainty-MGV]{0.24\linewidth}
\includegraphics[width=1.0\textwidth]{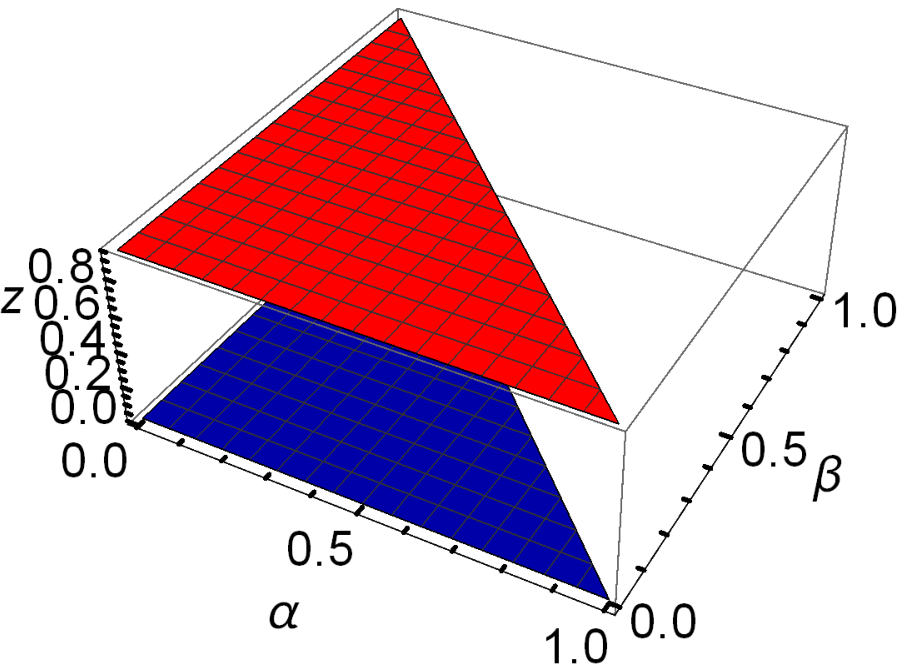}
\end{minipage}}
\subfigure[]
{\begin{minipage}[2-d-XuCong-Uncertainty-MGV]{0.24\linewidth}
\includegraphics[width=1.0\textwidth]{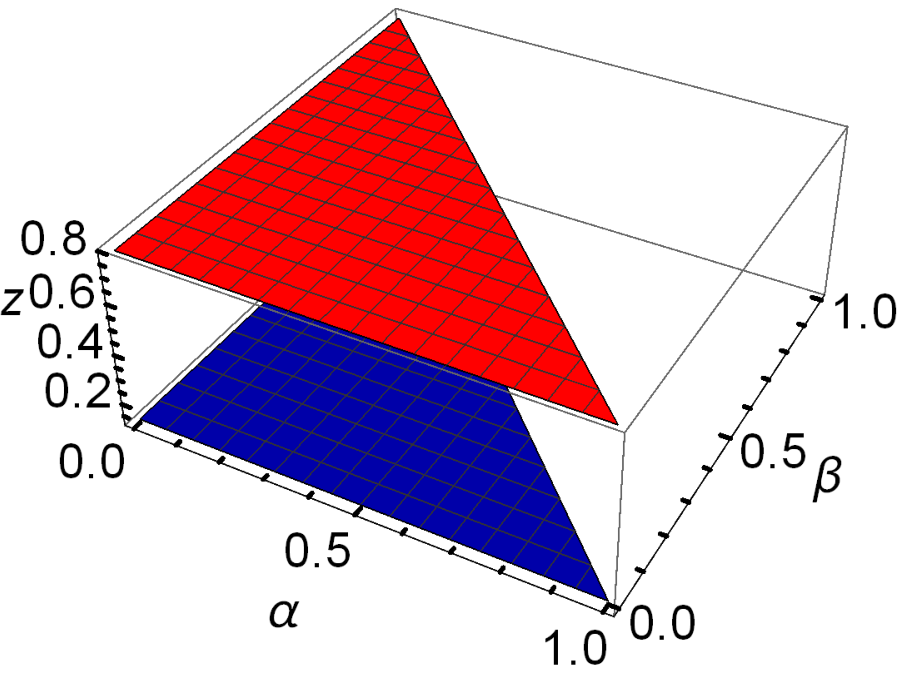}
\end{minipage}}
\subfigure[]
{\begin{minipage}[2-e-XuCong-Uncertainty-MGV]{0.24\linewidth}
\includegraphics[width=1.0\textwidth]{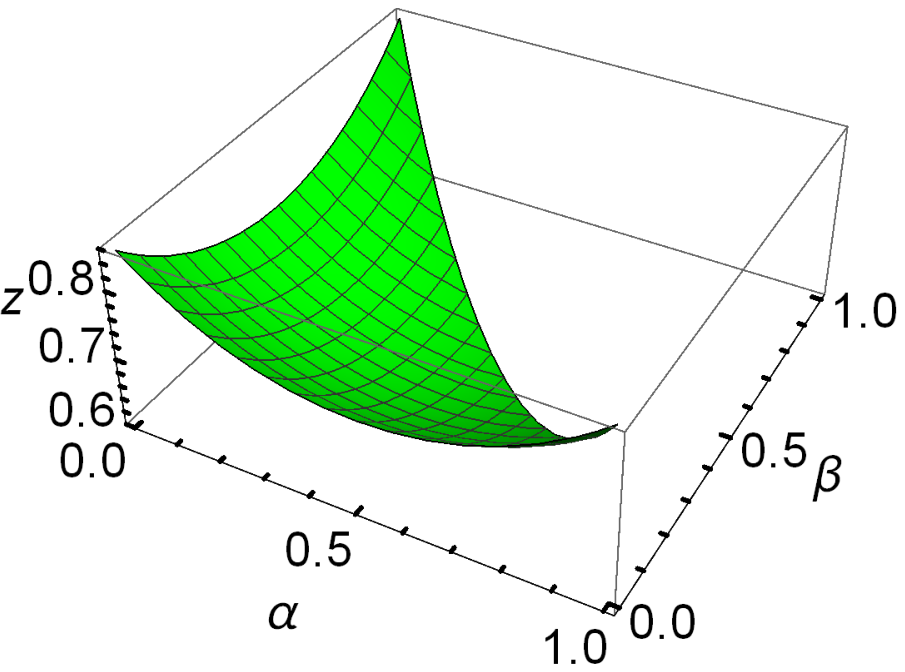}
\end{minipage}}
\subfigure[]
{\begin{minipage}[2-f-XuCong-Uncertainty-MGV]{0.24\linewidth}
\includegraphics[width=1.0\textwidth]{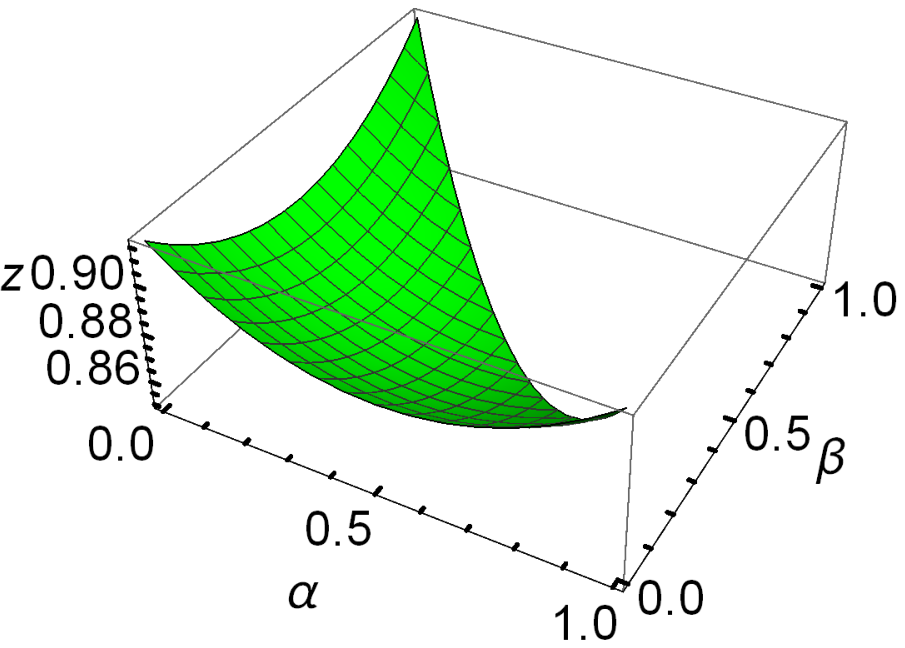}
\end{minipage}}
\subfigure[]
{\begin{minipage}[2-g-XuCong-Uncertainty-MGV]{0.24\linewidth}
\includegraphics[width=1.0\textwidth]{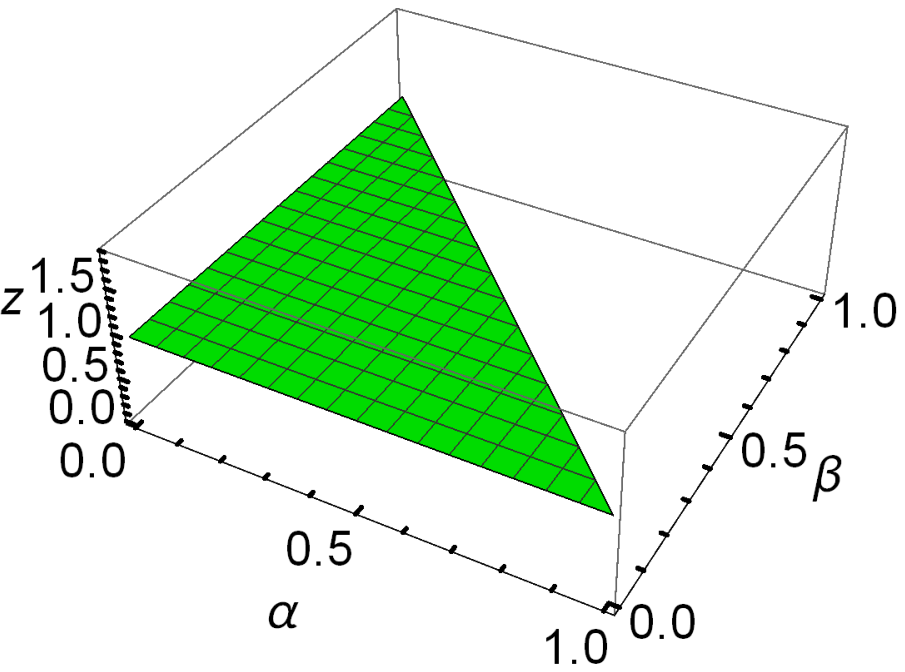}
\end{minipage}}
\subfigure[]
{\begin{minipage}[2-h-XuCong-Uncertainty-MGV]{0.24\linewidth}
\includegraphics[width=1.0\textwidth]{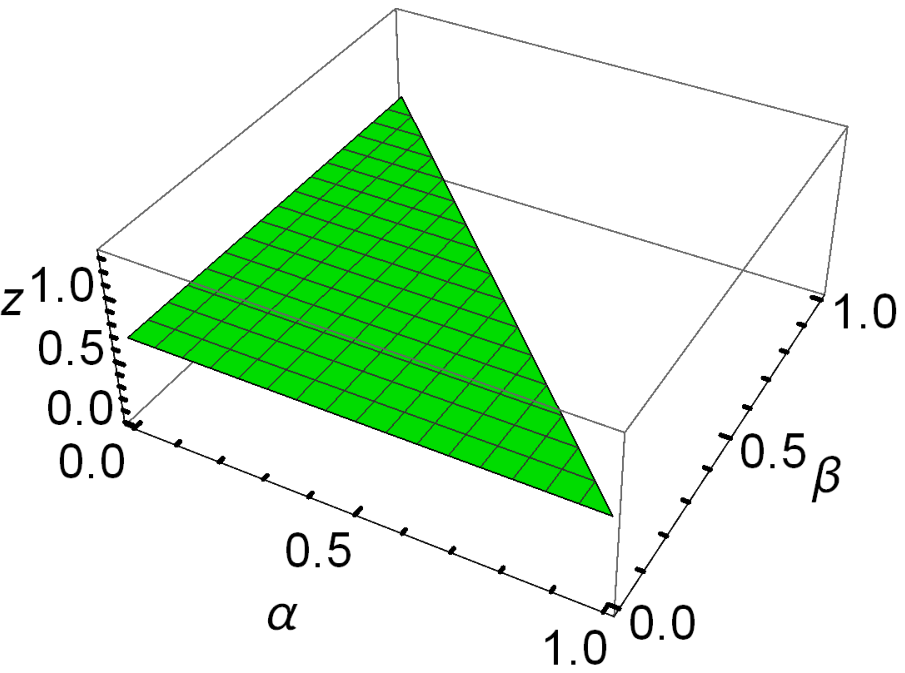}
\end{minipage}}
\caption{Surfaces of
$V^{\alpha,\beta}(\rho_w,\Phi)$ and $Q^{\alpha,\beta} (\rho_w,\Phi)$
with fixed $p$: $(\mathbf{a})$ $p=\frac{1}{4}$; $(\mathbf{b})$
$p=\frac{1}{2}$; $(\mathbf{c})$ $p=\frac{3}{4}$; $(\mathbf{d})$
$p=1$, where the red (blue) surface represents the value of
$V^{\alpha,\beta}(\rho_w,\Phi)$ ($Q^{\alpha,\beta} (\rho_w,\Phi)$)
in Eq. (\ref{eq31}) (Eq. (\ref{eq32})). Surfaces of the gap
$C^{\alpha,\beta}(\rho_w,\Phi)$ in $(\mathbf{a})$-$(\mathbf{d})$
with fixed $p$: $(\mathbf{e})$ $p=\frac{1}{4}$; $(\mathbf{f})$
$p=\frac{1}{2}$; $(\mathbf{g})$ $p=\frac{3}{4}$; $(\mathbf{h})$
$p=1$, where the green surface represents the value of
$C^{\alpha,\beta}(\rho_w,\Phi)$.}
   \label{fig:Fig2}
\end{figure}

Now consider the isotropic state,
$$\rho_{iso}=\left(
\begin{array}{cccc}
 \frac{1}{6} (2 F+1) & 0 & 0 & \frac{1}{6} (4 F-1) \\
 0 & \frac{1}{3}(1-F) & 0 & 0 \\
 0 & 0 & \frac{1}{3}(1-F) & 0 \\
 \frac{1}{6} (4 F-1) & 0 & 0 & \frac{1}{6} (2 F+1) \\
\end{array}
\right),
$$
where $F\in[0,1]$, which is separable when $F\in
[0,\frac{1}{2}]$. According to Eqs. (\ref{eq29}) and (\ref{eq30}), we obtain
\begin{eqnarray}\label{eq33}
V^{\alpha,\beta}(\rho_{iso},\Phi)
&=&\frac{1}{24}\left[19-2F-8F^2+3^{2-\alpha-\beta}F^{1-\alpha-\beta}(1-F)^{\alpha+\beta}\right.
\nonumber\\
&&\left.+3^{1+\alpha+\beta}F^{\alpha+\beta}(1-F)^{1-\alpha-\beta} \right]
\end{eqnarray}
and
\begin{eqnarray}\label{eq34}
&&Q^{\alpha,\beta}(\rho_{iso},\Phi)\nonumber\\
&=&\frac{1}{8}\left[1+2F-3^{1-\alpha}F^{1-\alpha}(1-F)^{\alpha}
-3^{\alpha}F^{\alpha}(1-F)^{1-\alpha}-3^{1-\beta}F^{1-\beta}(1-F)^{\beta}-\right.
\nonumber\\
&&\left.3^{\beta}F^{\beta}(1-F)^{1-\beta}+3^{\alpha+\beta}F^{\alpha
+\beta}(1-F)^{1-\alpha-\beta}+3^{1-\alpha-\beta}F^{1-\alpha-\beta}(1-F)^{\alpha+\beta}\right].
\end{eqnarray}
Fig.~\ref{fig:Fig3} illustrates the uncertainties
$V^{\alpha,\beta}(\rho_{iso},\Phi)$ and
$Q^{\alpha,\beta}(\rho_{iso},\Phi)$ in Eqs. (\ref{eq30}) and
(\ref{eq31}) with $\alpha=\frac{1}{5}$ and $\beta=\frac{3}{10}$,
respectively. By calculation, it is found that
$V^{\alpha,\beta}(\rho_{iso},\Phi)-Q^{\alpha,\beta}(\rho_{iso},\Phi)$
reaches its maximum value of $0.9375$ when $F=\frac{1}{4}$,
$\alpha=0.210985$ and $\beta=0.479723$. Interestingly, the linear
entropy (mixedness)
$1-\mathrm{tr}{\rho_{iso}^2}=\frac{2}{3}+\frac{2}{3}F-\frac{4}{3}F^2$
of the isotropic state attains its maximum value of $\frac{3}{4}$
when $F=\frac{1}{4}$. In Fig.~\ref{fig:Fig4}, we plot the surfaces
of the total uncertainty $V^{\alpha,\beta}(\rho_{iso},\Phi)$ and the
quantum uncertainty $Q^{\alpha,\beta}(\rho_{iso},\Phi)$ in Eqs.
(\ref{eq33}) and (\ref{eq34}), respectively, and their gap, the
classical uncertainty $C^{\alpha,\beta}(\rho_{iso},\Phi)$, for fixed
values of $F$.
\begin{figure}[ht]\centering
{\begin{minipage}[XuCong-Uncertainty-MGV-3]{0.5\linewidth}
\includegraphics[width=0.95\textwidth]{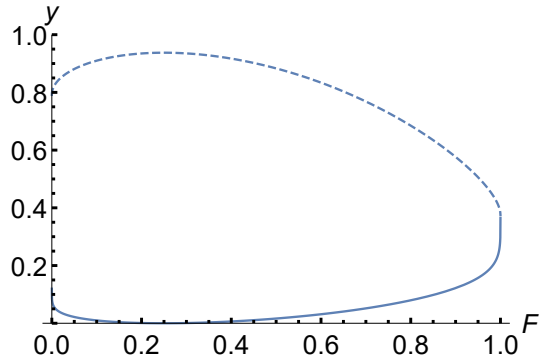}
\end{minipage}}
\caption{{The $y$-axis shows the values of
$V^{\alpha,\beta}(\rho_{iso},\Phi)$ and $Q^{\alpha,\beta}
(\rho_{iso},\Phi)$. Dashed (solid) line represents the value of
$V^{\alpha,\beta}(\rho_{iso},\Phi)$
($Q^{\alpha,\beta}(\rho_{iso},\Phi)$) in Eq. (\ref{eq33}) (Eq.
(\ref{eq34})) with $\alpha=\frac{1}{5}$ and $\beta=\frac{3}{10}$,
respectively. The isotropic state $\rho_{iso}$ degenerates into a
pure state when $F=1$ and in this case
$V^{\alpha,\beta}(\rho_{iso},\Phi)=Q^{\alpha,\beta}
(\rho_{iso},\Phi)$.\label{fig:Fig3}}}
\end{figure}

\begin{figure}[H]\centering
\subfigure[]
{\begin{minipage}[4-a-XuCong-Uncertainty-MGV]{0.24\linewidth}
\includegraphics[width=1.0\textwidth]{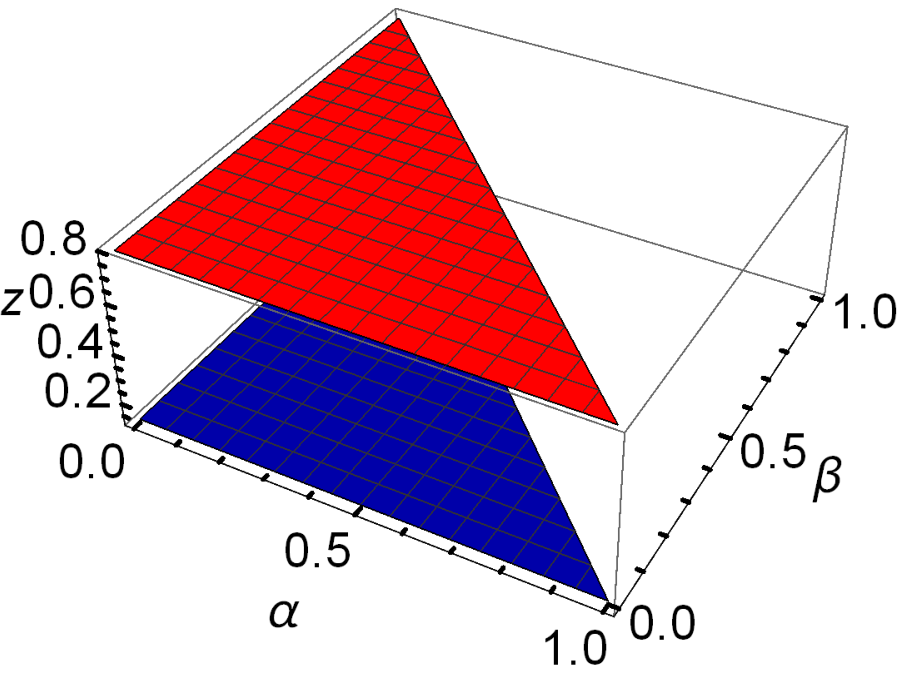}
\end{minipage}}
\subfigure[]
{\begin{minipage}[4-b-XuCong-Uncertainty-MGV]{0.24\linewidth}
\includegraphics[width=1.0\textwidth]{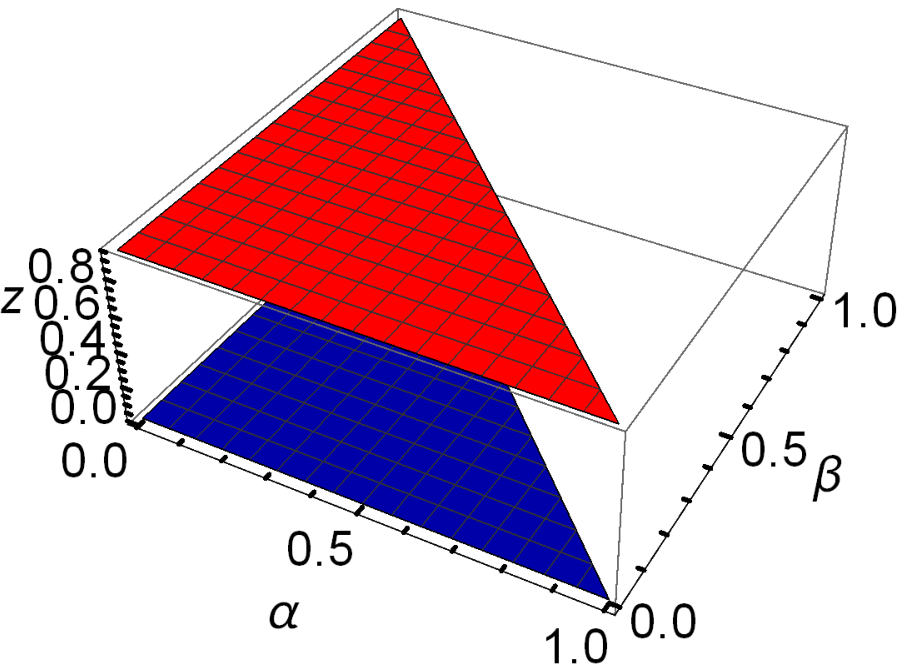}
\end{minipage}}
\subfigure[]
{\begin{minipage}[4-c-XuCong-Uncertainty-MGV]{0.24\linewidth}
\includegraphics[width=1.0\textwidth]{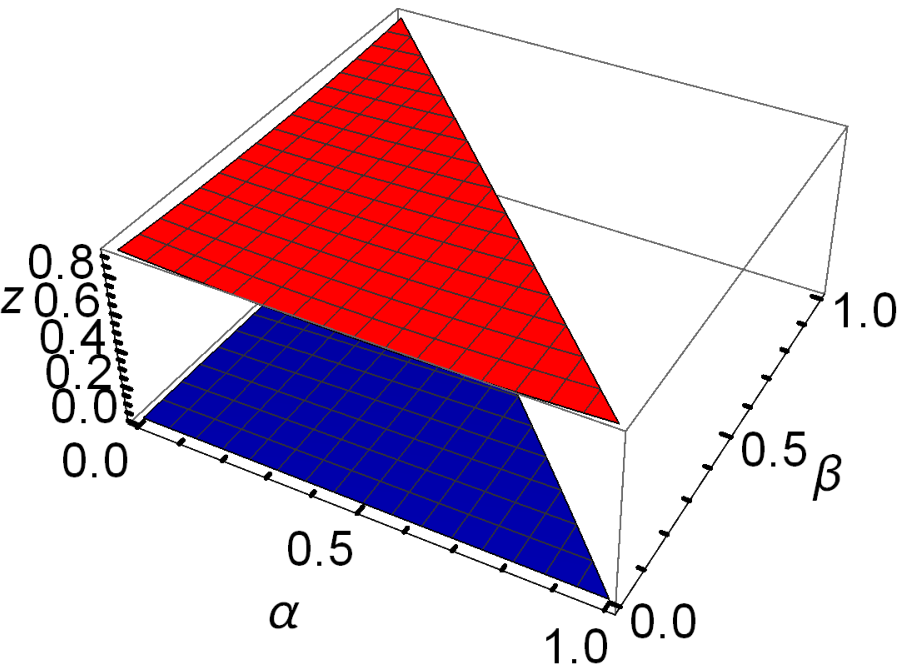}
\end{minipage}}
\subfigure[]
{\begin{minipage}[4-d-XuCong-Uncertainty-MGV]{0.24\linewidth}
\includegraphics[width=1.0\textwidth]{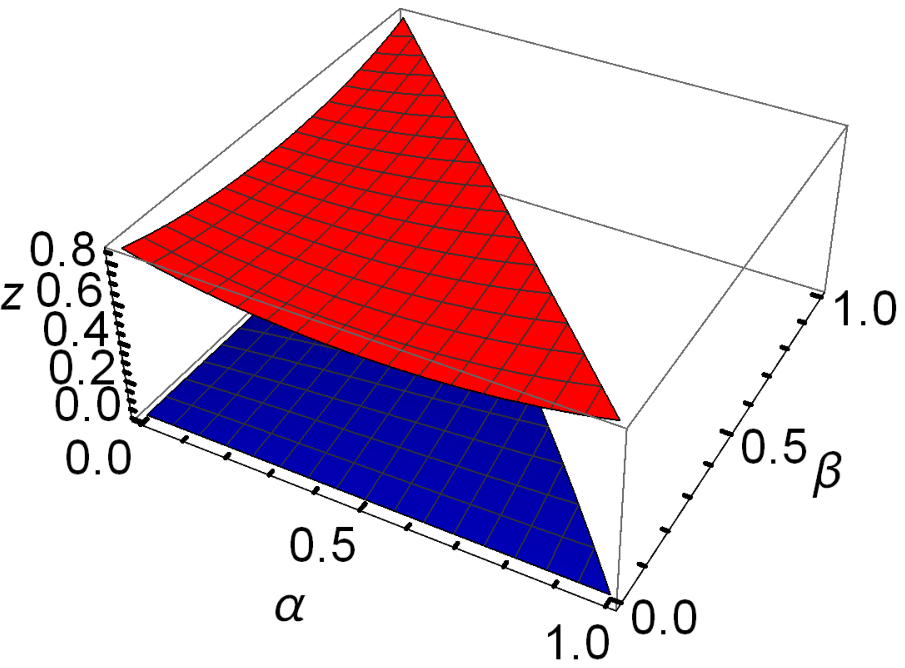}
\end{minipage}}
\subfigure[]
{\begin{minipage}[4-e-XuCong-Uncertainty-MGV]{0.24\linewidth}
\includegraphics[width=1.0\textwidth]{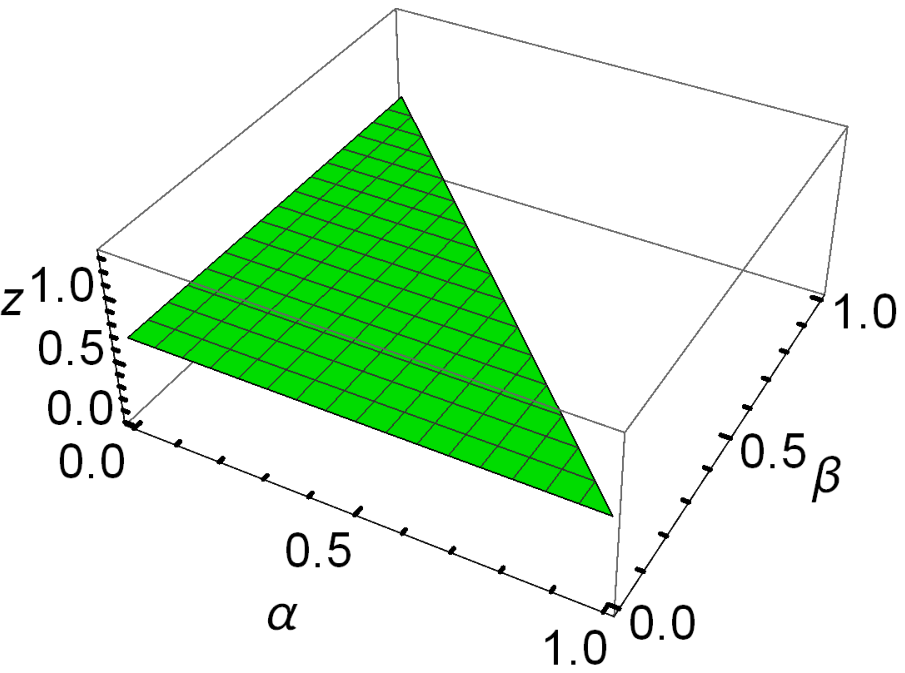}
\end{minipage}}
\subfigure[]
{\begin{minipage}[4-f-XuCong-Uncertainty-MGV]{0.24\linewidth}
\includegraphics[width=1.0\textwidth]{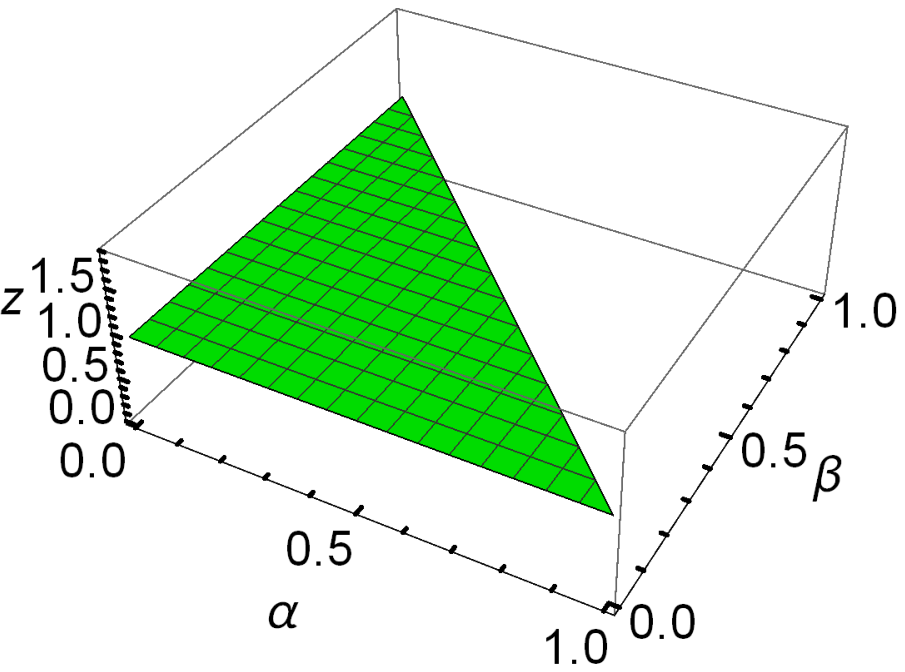}
\end{minipage}}
\subfigure[]
{\begin{minipage}[4(g)XuCong-Uncertainty-MGV]{0.24\linewidth}
\includegraphics[width=1.0\textwidth]{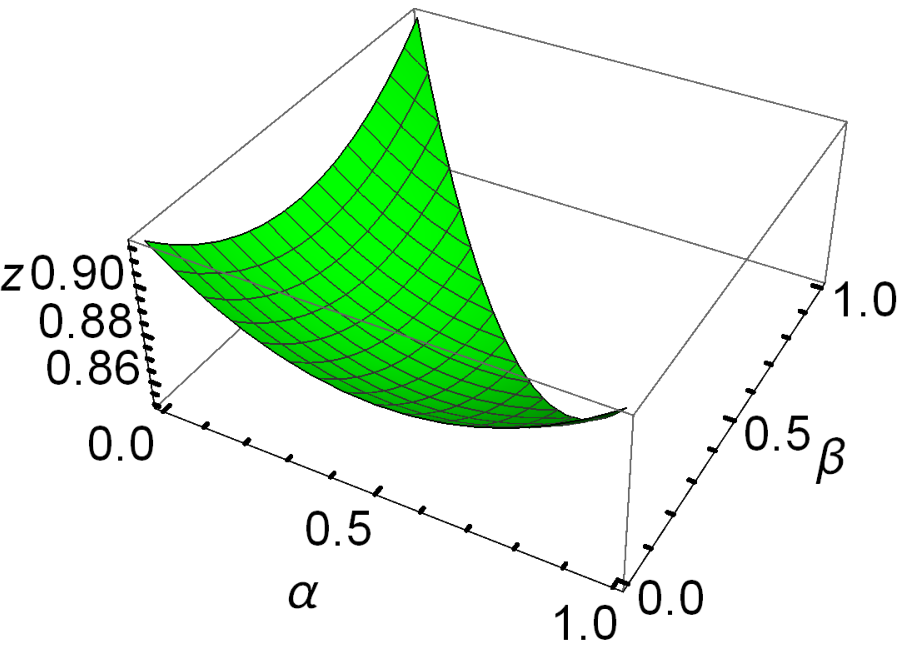}
\end{minipage}}
\subfigure[]
{\begin{minipage}[4-h-XuCong-Uncertainty-MGV]{0.24\linewidth}
\includegraphics[width=1.0\textwidth]{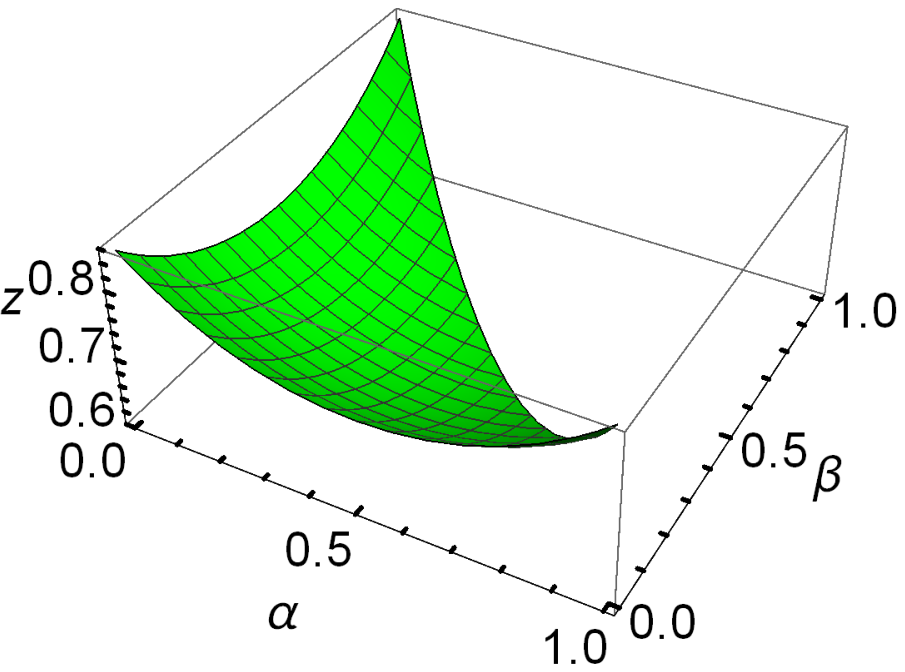}
\end{minipage}}
\caption{Surfaces of
$V^{\alpha,\beta}(\rho_{iso},\Phi)$ and $Q^{\alpha,\beta}
(\rho_{iso},\Phi)$ with fixed $F$: $(\mathbf{a})$ $F=0$;
$(\mathbf{b})$ $F=\frac{1}{4}$; $(\mathbf{c})$ $F=\frac{1}{2}$;
$(\mathbf{d})$ $F=\frac{3}{4}$, where the red (blue) surface
represents the value of $V^{\alpha,\beta}(\rho_{iso},\Phi)$
($Q^{\alpha,\beta} (\rho_{iso},\Phi)$) in Eq. (\ref{eq33}) (Eq.
(\ref{eq34})). Surfaces of the gap
$C^{\alpha,\beta}(\rho_{iso},\Phi)$ in $(\mathbf{a})$-$(\mathbf{d})$
with fixed $F$: $(\mathbf{e})$ $F=0$; $(\mathbf{f})$
$F=\frac{1}{4}$; $(\mathbf{g})$ $F=\frac{1}{2}$; $(\mathbf{h})$
$F=\frac{3}{4}$, where the green surface represents the value of
$C^{\alpha,\beta}(\rho_{iso},\Phi)$.}
\label{fig:Fig4}
\end{figure}

By plotting the animation of Fig.~\ref{fig:Fig2}
and Fig.~\ref{fig:Fig4} for $p\in[0,1]$ and $F\in[0,1]$,
respectively, it can be seen that for the Werner state $\rho_w$, the
maximum value of classical uncertainty of the channel
$C^{\alpha,\beta}(\rho_w,\Phi)$ increases with the increase of $p$
when $p\in[0,\frac{3}{4}]$, and decreases with the increase of $p$
when $p\in[\frac{3}{4},1]$, while for the isotropic state
$\rho_{iso}$, the maximum value of
$C^{\alpha,\beta}(\rho_{iso},\Phi)$ increases with the increase of
$F$ when $F\in[0,\frac{1}{4}]$, and decreases with the increase of
$F$ when $F\in[\frac{1}{4},1]$. For either the Werner state $\rho_w$
or the isotropic state $\rho_{iso}$, the classical uncertainty of
the channel is maximal when the linear entropy (mixedness) is
maximal.

\noindent{\bf Example 6} Consider the von Neumann
measurement $\Pi(\rho)=\sum_i\Pi_i\rho\Pi_i$, where
$\Pi_i=|i\rangle\langle i|$ with ${\{|i\rangle\}_{i=1}^d}$ being an
orthonormal basis of $\mathcal{H}$. It is easy to derive that
\begin{equation}\label{eq35}
V^{\alpha,\beta}(\rho,\Pi)=\frac{1}{2}\left[1+\sum_i \langle
i|{\rho}^{\alpha+\beta}|i\rangle\langle
i|{\rho}^{1-\alpha-\beta}|i\rangle\right]-\sum_i{\langle
i|\rho|i\rangle}^2
\end{equation}
and
\begin{eqnarray}\label{eq36}
Q^{\alpha,\beta}(\rho,\Pi) &=&\frac{1}{2}\left[1+\sum_i{\langle
i|{\rho}^{\alpha+\beta}|i\rangle}{\langle
i|{\rho}^{1-\alpha-\beta}|i\rangle}-\sum_i{\langle
i|{\rho}^{\beta}|i\rangle}{\langle
i|{\rho}^{1-\beta}|i\rangle}\right.
\nonumber\\
&&\left.-\sum_i{\langle i|{\rho}^{1-\alpha}|i\rangle}{\langle
i|{\rho}^{\alpha}|i\rangle}\right],
\end{eqnarray}
where $\alpha,\beta\geq 0$ and $\alpha+\beta\leq 1$.
\vskip0.1in

Now we use Example 6 to illustrate Eqs.
(\ref{eq18}), (\ref{eq19}) and (\ref{eq20}) for two classes of
states, the Werner state and the isotropic state, respectively.
Under this channel, the left hand side of
inequality (\ref{eq18}) is $\frac{1}{2}\left(1+\sum_i \langle
i|{\rho}^{\alpha+\beta}|i\rangle\langle
i|{\rho}^{1-\alpha-\beta}|i\rangle\right)$, which is equal to
$\frac{3}{4}+\frac{p}{6}+\frac{1}{4}[3^{\alpha+\beta-1}p^{1-\alpha-\beta}(1-p)^{\alpha+\beta}+3^{-\alpha-\beta}p^{\alpha+\beta}(1-p)^{1-\alpha-\beta}]$
for the Werner state $\rho_w$. The right hand side of (\ref{eq18}),
however, is $\frac{1}{2}(1+\mathrm{max}_i\rho^
{1-\alpha-\beta}_{ii}\mathrm{tr}\rho^{\alpha+\beta})$, which is
equal to
$\frac{3}{4}+\frac{1}{4}(3^{1-\alpha-\beta}p^{\alpha+\beta}(1-p)^{1-\alpha-\beta}+3^{\alpha+\beta-1}p^{1-\alpha-\beta}(1-p)^{\alpha+\beta})$
when $p\in[0,\frac{3}{4})$, and
$\frac{p}{2}+\frac{1}{2}(1+3^{\alpha+\beta-1}p^{1-\alpha-\beta}(1-p)^{\alpha+\beta})$
when $p\in(\frac{3}{4},1]$ for the Werner state $\rho_w$.
When $p=\frac{3}{4}$, the left and right hand side
of (\ref{eq18}) are both $1$. For the isotropic state $\rho_{iso}$,
the left hand side of (\ref{eq18}) turns out to be
$\frac{11}{12}-\frac{F}{6}+\frac{1}{4}[3^{-\alpha-\beta}F^{1-\alpha-\beta}(1-F)^{\alpha+\beta}+3^{\alpha+\beta-1}F^{\alpha+\beta}(1-F)^{1-\alpha-\beta}]$,
while the right hand side is
$1+\frac{1}{2}(3^{\alpha+\beta-1}F^{\alpha+\beta}(1-F)^{1-\alpha-\beta}-F)$
when $F\in[0,\frac{1}{4})$, and
$\frac{3}{4}+\frac{1}{4}(3^{1-\alpha-\beta}F^{1-\alpha-\beta}(1-F)^{\alpha+\beta}+3^{\alpha+\beta-1}F^{\alpha+\beta}(1-F)^{1-\alpha-\beta})$
when $F\in(\frac{1}{4},1]$. When $F=\frac{1}{4}$,
the left and right hand side of (\ref{eq18}) are both $1$.

The left and right hand side of Eq. (\ref{eq19}) for the Werner
state $\rho_w$ are
\begin{equation}\label{eq37}
1+(2V^{\alpha,\beta}(\rho_w,\Phi)+1-\mathrm{tr}\rho_w^{\alpha+\beta}\Phi
(\rho_w^{1-\alpha-\beta}))\mathrm{log}4=3+\frac{8}{3}p-\frac{16}{3}p^2
\end{equation}
and
\begin{equation}\label{eq38}
S_e(\rho_w,\Phi)=(p-1)\log(1-p)-p\log\frac{p}{3},
\end{equation}
respectively, while the left and right hand side of Eq. (\ref{eq20})
are given by
\begin{eqnarray}\label{eq39}
&&2(2V^{\alpha,\beta}(\rho_w,\Phi)+1-\mathrm{tr}\rho_w^{\alpha+\beta}\Phi (\rho_w^{1-\alpha-\beta}))\mathrm{log}4+I_c({\rho_w},\Phi)\nonumber \\
&=&4+\frac{16}{3}p-\frac{32}{9}p^2+(1-p)\log(1-p)+\frac{p}{3}\log\frac{p}{3}-\frac{3-2p}{3}\log\frac{3-2p}{6}
\end{eqnarray}
and
 \begin{equation}\label{eq40}
 S(\rho_w)-2=(p-1)\log(1-p)-p\log\frac{p}{3}-2,
 \end{equation}
respectively. Fig.~\ref{fig:Fig5} illustrates the quantities in Eqs.
(\ref{eq37})-(\ref{eq40}) for the Werner state $\rho_w$ as a
function of $p$.
\begin{figure}[ht]\centering
{\begin{minipage}[XuCong-Uncertainty-MGV-5]{0.5\linewidth}
\includegraphics[width=0.95\textwidth]{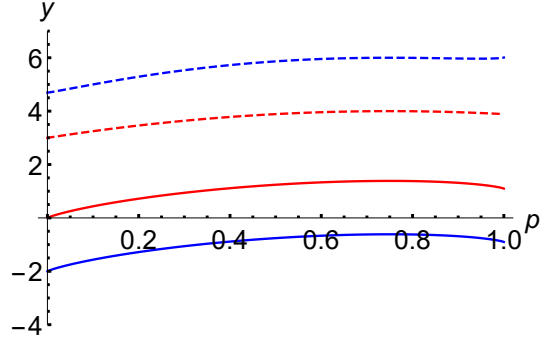}
\end{minipage}}
\caption{{The $y$-axis shows the uncertainty and its lower bounds.
Dashed red (solid red) line represents the value of Eq. (\ref{eq37})
(Eq. (\ref{eq38})) for the Werner state $\rho_w$; Dashed blue (solid
blue) line represents the value of Eq. (\ref{eq39}) (Eq.
(\ref{eq40})) for the Werner state $\rho_w$. \label{fig:Fig5}}}
\end{figure}

Now consider the isotropic state $\rho_{iso}$. Direct computation
shows that the quantities on the two sides of Eq. (\ref{eq19}) are
\begin{equation}\label{eq41}
1+(2V^{\alpha,\beta}(\rho_{iso},\Phi)+1-\mathrm{tr}\rho_{iso}^{\alpha+\beta}\Phi
(\rho_{iso}^{1-\alpha-\beta}))\mathrm{log}4=\frac{35}{9}+\frac{8}{9}F-\frac{16}{9}F^2
\end{equation}
and
\begin{equation}\label{eq42}
S_e(\rho_{iso},\Phi)=(F-1)\log\frac{1-F}{3}-F\log F,
\end{equation}
respectively, and for Eq. (\ref{eq20}), we have
\begin{eqnarray}\label{eq43}
&&2(2V^{\alpha,\beta}(\rho_{iso},\Phi)+1-\mathrm{tr}\rho_{iso}^{\alpha+\beta}\Phi(\rho_{iso}^{1-\alpha-\beta}))\mathrm{log}4+I_c({\rho_{iso}},\Phi)\nonumber \\
&=&\frac{52}{9}+\frac{16}{9}F(1-2F)+\frac{1-F}{3}\log\frac{1-F}{3}+F\log
F-\frac{1+2F}{3}\log\frac{1+2F}{6}
\end{eqnarray}
and
 \begin{equation}\label{eq44}
 S(\rho_{iso})-2=(F-1)\log\frac{1-F}{3}-F\log F-2.
 \end{equation}
Fig.~\ref{fig:Fig6} illustrates the quantities in Eqs.
(\ref{eq41})-(\ref{eq44}) for the isotropic state $\rho_{iso}$ as a
function of $F$.
\begin{figure}[ht]\centering
{\begin{minipage}[XuCong-Uncertainty-MGV-6]{0.5\linewidth}
\includegraphics[width=0.95\textwidth]{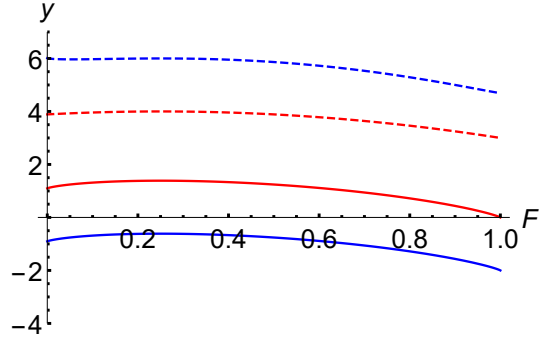}
\end{minipage}}
\caption{{The $y$-axis shows the uncertainty and its lower bounds.
Dashed red (solid red) line represents the value of Eq. (\ref{eq41})
(Eq. (\ref{eq42})) for the isotropic state $\rho_{iso}$; Dashed blue
(solid blue) line represents the value of Eq. (\ref{eq43}) (Eq.
(\ref{eq44})) for the isotropic state $\rho_{iso}$.
\label{fig:Fig6}}}
\end{figure}
Comparing Fig.~\ref{fig:Fig5} with
Fig.~\ref{fig:Fig6}, it can be seen that for the Werner state
$\rho_w$ (the isotropic state $\rho_{iso}$), the gap between Eq.
(\ref{eq39}) (Eq.(\ref{eq43})) and Eq. (\ref{eq40})
(Eq.(\ref{eq44})) is greater than the one between Eq. (\ref{eq37})
(Eq.(\ref{eq41})) and Eq. (\ref{eq38}) (Eq.(\ref{eq42})).

\vskip0.1in

Metric-adjusted skew information is known to be measurable by
measuring the linear response function at thermal equilibrium
\cite{STUM}. In recent years, uncertainty relations for unitary
operators have been investigated both theoretically and
experimentally \cite{BS,THSN,BKTN}, and the problem of measuring
arbitrary non-Hermitian operators has attracted much attention
\cite{PAKSSU,NGS,ZXZC}. We discuss how to measure the total/quantum
uncertainty of quantum channels
$V^{\alpha,\beta}({\rho},\Phi)$/$Q^{\alpha,\beta}({\rho},\Phi)$ for
pure states by using an experimental protocol.

We now focus on the case in which $\rho$ is a pure state, i.e.,
$\rho=|\psi\rangle\langle \psi|$. By using Eq. (\ref{eq13}), the
problem then reduces to how to measure the magnitude of the
expectation values of the non-Hermitian operators $K_i$.

An arbitrary operator $K\in\mathcal{B(H)}$ can be decomposed as
$K=UR$, where $U$ is unitary and $R=\sqrt{K^\dag K}$ is positive
semidefinite. By utilizing the scheme proposed in \cite{NGS}, we can
measure the magnitude of the expectation value $|\langle K\rangle|$
of a non-Hermitian operator $K$ by using the Mach-Zehnder
interferometer.

Given a quantum channel $\Phi$ with Kraus operators $\{K_i\}$. Using
the above scheme in \cite{NGS}, we can fix the values of $|\langle
K_i\rangle|$ for each $i$, and by Eq.(\ref{eq13}), both of
$V^{\alpha,\beta}({\rho},\Phi)$ and $Q^{\alpha,\beta}({\rho},\Phi)$
can be obtained as $\frac{1}{2}(1-\sum_i|\langle K_i\rangle|^2)$.

\vskip0.1in

\noindent {\bf 5. Conclusions and discussions}\\\hspace*{\fill}\\
We have defined a family of total uncertainties of quantum channels
based on the modified generalized  variance (MGV) introduced in
\cite{WZWLF}, which includes the total uncertainty introduced in
\cite{SL} as a special case. Following the idea in Ref.\cite{SL}, we
have divided the total uncertainty into quantum and classical parts
and associated the quantum part with the coherence of quantum states
with respect to quantum channels based on modified generalized
Wigner-Yanase-Dyson (MGWYD) skew information \cite{WZFL}.

In addition, we have formulated the relationship between the
MGV-based total uncertainty of quantum channels and the entanglement
fidelity, extending the results in \cite{SL} to a more general case.
As a consequence, we have also figured out the link between the
MGV-based total uncertainty of quantum channels and the entropy
exchange/coherent information, proving that the MGV-based total
uncertainty of quantum channels provides an upper bound on the
entropy exchange. Moreover, we have calculated the total uncertainty
and quantum uncertainty of some special quantum channels. In
particular, we have computed the total uncertainty and quantum
uncertainty for a quantum channel with respect to the Werner state
and the isotropic state in Example 5. It has been found that for
these two different classes of states, there exist subtle
similarity, that is, under this quantum channel, both the classical
uncertainty and the linear entropy (mixedness) attain their maximum
values for the same state, i.e., at the same value of the parameter
$p$ or $F$. The uncertainty relations Eqs. (\ref{eq19}) and
(\ref{eq20}) have also been computed for this channel and such two
classes of states. Finally, we have used an experimental protocol
formulated in \cite{NGS} to measure the uncertainty of quantum
channels for pure states. Our results may shed some new light on the
study of uncertainties of quantum channels and provide new insights
into the quantum-classical interplay.

\vskip0.1in

\noindent

\subsubsection*{Acknowledgements}
\small {The authors would like to thank the anonymous referees for
their valuable comments and suggestions, which have greatly improved
this paper. They would also like to thank Dr. Lin Zhang for fruitful
discussions. Zhaoqi Wu was supported by National Natural Science
Foundation of China (Grant Nos. 12161056, 11701259); Jiangxi
Provincial Natural Science Foundation (Grant No. 20202BAB201001);
Shao-Ming Fei was supported by National Natural Science Foundation
of China (Grant Nos. 12075159, 12171044); Beijing Natural Science
Foundation (Grant No. Z190005); Academy for Multidisciplinary
Studies, Capital Normal University; Shenzhen Institute for Quantum
Science and Engineering, Southern University of Science and
Technology (Grant No. SIQSE202001); the Academician Innovation
Platform of Hainan Province.}


\subsubsection*{Competing interests}
\small {The authors declare no competing interests.}


\subsubsection*{Data availability}
\small {Data sharing not applicable to this article as no datasets
were generated or analysed during the current study.}



\begin{thebibliography}{S2}

\bibitem{HEISENBERG} Heisenberg, W.: \"uber den anschaulichen Inhalt der
quantentheoretischen Kinematik und Mechanik. Z. Phys. \textbf{43},
172 (1927)

\bibitem{ROBERT} Robertson, H.P.: The uncertainty principle. Phys. Rev. \textbf{34}, 163 (1929)

\bibitem{LUO1} Luo, S.: Heisenberg uncertainty relation for mixed states. Phys. Rev. A \textbf{72}, 042110 (2005)


\bibitem{LUO2} Luo, S.: Quantum versus classical uncertainty. Theor. Math. Phys. \textbf{143}, 681 (2005)

\bibitem{LUO3} Luo, S.: Quantum uncertainty of mixed states based on skew information. Phys. Rev. A \textbf{73}, 022324
(2006)
\bibitem{GUDDER} Gudder, S.: Operator probability theory. Int. J. Pure Appl. Math. \textbf{39}, 511 (2007)

\bibitem{DD1}Dou, Y., Du, H.: Generalizations of the Heisenberg and Schr\"{o}dinger uncertainty relations. J. Math.
Phys. \textbf{54}, 103508 (2013)

\bibitem{DD2} Dou, Y., Du, H.: Note on the Wigner-Yanase-Dyson skew information. Int. J. Theor. Phys. \textbf{53}, 952 (2014)

\bibitem{SL} Sun, Y., Li, N.: The uncertainty of quantum channels in terms of variance. Quantum Inf. Process. \textbf{20}, 25 (2021)

\bibitem{WY} Wigner, E.P., Yanase, M.M.: Information contents of distributions. Proc. Natl. Acad. Sci. USA \textbf{49}, 910 (1963)
\bibitem{CL} Chen, P., Luo, S.: Direct approach to quantum extensions of Fisher information. Front. Math. China \textbf{2}, 359 (2007)
\bibitem{YK1} Yanagi, K.: Uncertainty relation on Wigner-Yanase-Dyson skew information. J. Math. Anal. Appl. \textbf{365}, 12 (2010)
\bibitem{YK2} Yanagi, K.: Wigner-Yanase-Dyson skew information and uncertainty relation. J. Phys. Conf. Ser. \textbf{201}, 012015 (2010)
\bibitem{CAIL}Cai, L., Luo, S.: On convexity of generalized Wigner-Yanase-Dyson information. Lett. Math. Phys. \textbf{83}, 253 (2008)

\bibitem{MN} Moiseyev, N.: Non-Hermitian Quantum Mechanics. Cambridge University Press, Cambridge (2011)
\bibitem{NC} Nielson, M.A., Chuang, I.L.: Quanutm Computation and Quantum Information (10th Anniversary Edition). Cambridge University Press, Cambridge (2010)
\bibitem{LGL} Long, G.L.: General quantum interference principle and duality computer. Commun. Theor. Phys. \textbf{45}, 825 (2006)
\bibitem{BBS} Bender, C.M., Boettcher, S.: Real spectra in non-Hermitian Hamiltonians having PT Symmetry. Phys. Rev. Lett. \textbf{80}, 5243 (1998)
\bibitem{MKGE} Makris, K.G., El-Ganainy, R., Christodoulides, D., Musslimani, Z.H.: Beam dynamics in PT Symmetric optical lattices. Phys. Rev. Lett. \textbf{100}, 103904 (2008)
\bibitem{GSDM} Guo, A., Salamo, G., Duchesne, D., Morandotti, R., Volatier-Ravat, M., Aimez, V., Siviloglou, G., Christodoulides, D.: Observation of PT-symmetry breaking in complex optical potentials. Phys. Rev. Lett. \textbf{103}, 093902 (2009)
\bibitem{RMER} Ruter, C.E., Makris, K.G., El-Ganainy, R., Christodoulides, D.N., Segev, M., Kip, D.: Observation of parity-time symmetry in optics. Nat. Phys. \textbf{6}, 192 (2010)
\bibitem{CJHY} Chang, L., Jiang, X., Hua, S., Yang, C., Wen, J., Jiang, L., Li, G., Wang, G., Xiao, M.: Parity-time symmetry and variable optical isolation in active-passive-coupled microresonators. Nat. Photon. \textbf{8}, 524 (2014)
\bibitem{TWYH} Tang, J.-S., Wang, Y.-T., Yu, S., He, D.-Y., Xu, J.-S., Liu, B.-H., Chen, G., Sun, Y.-N., Sun, K., Han, Y.-J., Li, C.-F., Guo, G.-C.: Experimental investigation of the no-signalling principle in parity-time symmetric theory using an open quantum system. Nat. Photon. \textbf{10}, 642 (2016)

\bibitem{WZWLF} Wu, Z., Zhang, L., Wang, J., Li-Jost, X., Fei, S.-M.: Uncertainty relations based on modified Wigner-Yanase-Dyson skew information. Int. J. Theor. Phys. \textbf{59}, 704 (2020)
\bibitem{WZFL} Wu, Z., Zhang, L., Fei, S.-M., Li-Jost, X.: Coherence and complementarity based on modified generalized skew information. Quantum Inf. Process. \textbf{19}, 154 (2020)
\bibitem{BG} Busch, P., Grabowski, M., Lahti, P.: Operational Quantum Physics, 2nd edn. Springer, Berlin (1997)



\bibitem{AD} Ahn, D.: Unruh effect as a noisy quantum channel. Phys. Rev. A \textbf{98}, 022308 (2018)
\bibitem{CQYP} Chen, N., Quan, D., Yang, H., Pei, C.: Deterministic controlled remote state preparation using partially entangled quantum channel. Quantum Inf. Process. \textbf{15}, 1719 (2016)
\bibitem{FF} Fang, K., Fawzi, H.: Geometric R\'enyi divergence and its applications in quantum channel capacities. Commun. Math. Phys. \textbf{384}, 1615 (2021)
\bibitem{YFMF} Yang, Z., Fan, Z., Mu, L., Fan, H.: Approximate quantum state reconstruction without a quantum channel. Phys. Rev. A \textbf{98}, 062315 (2018)
\bibitem{MS} Macchiavello, C., Sacchi, M.F.: Detecting lower bounds to quantum channel capacities. Phys. Rev. Lett. \textbf{116}, 140501 (2016)
\bibitem{ZHA} Zadeh, M.S.S., Houshmand, M., Aghababa, H.: Bidirectional teleportation of a two-qubit state by using eight-qubit entangled state as a quantum channel. Int. J. Theor. Phys. \textbf{56}, 2101 (2017)
\bibitem{HSTR} Ho, M., Sekatski, P., Tan, E.Y.Z., Renner, R., Bancal, J.D., Sangouard, N.: Noisy preprocessing facilitates a photonic realization of device-independent quantum key distribution. Phys. Rev. Lett. \textbf{124}, 230502 (2020)
\bibitem{MR1} Mabena, C.M., Roux, F.S.: High-dimensional quantum channel estimation using classical light. Phys. Rev. A \textbf{96}, 053860 (2017)
\bibitem{MK}Mani, A.,  Karimipour, V.: Cohering and decohering power of quantum channels. Phys. Rev. A \textbf{92}, 032331 (2015)
\bibitem{LZZ} Lv, S.-X., Zhao, Z.-W., Zhou, P.: Joint remote control of an arbitrary single-qubit state by using a multiparticle entangled state as the quantum channel. Quantum Inf. Process. \textbf{17}, 8 (2018)
\bibitem{MR2} Mabena, C.M., Roux, F.S.: Quantum channel correction with twisted light using compressive sensing. Phys. Rev. A \textbf{101}, 013807 (2020)
\bibitem{MM} Memarzadeh, L., Mancini, S.: Minimum output entropy of a non-Gaussian quantum channel. Phys. Rev. A \textbf{94}, 022341 (2016)
\bibitem{TFIY} Takeuchi, Y., Fujii, K., Ikuta, R., Yamamoto, T., Imoto, N.: Blind quantum computation over a collective-noise channel. Phys. Rev. A \textbf{93}, 052307 (2016)
\bibitem{WILDE} Wilde, M.M.: Entanglement cost and quantum channel simulation. Phys. Rev. A \textbf{98}, 042338 (2018)
\bibitem{BGD} Bej, P., Ghosal, A., Das, D., Roy, A., Bandyopadhyay, S.: Information-disturbance trade-off in generalized entanglement swapping. Phys. Rev. A \textbf{102}, 052416 (2020)
\bibitem{BFI} Buscemi, F., Sacchi, M.F.: Information-disturbance trade-off in quantum-state discrimination. Phys. Rev. A \textbf{74}, 052320 (2006)
\bibitem{SST} Shiraishi, N., Saito, K., Tasaki, H.: Universal trade-off relation between power and efficiency for heat engines. Phys. Rev. Lett. \textbf{117}, 190601 (2016)
\bibitem{Luo4} Luo, S.: Information conservation ang entropy change in quantum measurements. Phys. Rev. A \textbf{82}, 052103 (2010)
\bibitem{KJR} Korzekwa, K., Jennings, D., Rudolph, T.: Operational constraints on state-dependent formulations of quantum error-disturbance trade-off relations. Phys. Rev. A \textbf{89}, 052108 (2014)
\bibitem{MAS} Mandayam, P., Srinivas, M.D.: Disturbance trade-off principle for quantum measurements. Phys. Rev. A \textbf{90}, 062128 (2014)
\bibitem{KJJY} Kwon, H., Jeong, H., Jennings, D., Yadin, B., Kim, M.S.: Clock-work trade-off relation for coherence in quantum thermodynamics. Phys. Rev. Lett. \textbf{120}, 150602 (2018)
\bibitem{SP} Sharma, G., Pati, A.K.: Trade-off relation for coherence and disturbance. Phys. Rev. A \textbf{97}, 062308 (2018)
\bibitem{SKU} Shitara, T., Kuramochi, Y., Ueda, M.: Trade-off relation between information and disturbance in quantum measurement. Phys. Rev. A \textbf{93}, 032134 (2016)
\bibitem{PS} Pietzonka, P., Seifert, U.: Universal trade-off between power, efficiency, and constancy in steady-state heat engines. Phys. Rev. Lett. \textbf{120}, 190602 (2018)
\bibitem{SB1} Schumacher, B.: Sending entanglement through noisy quantum channels. Phys. Rev. A \textbf{54}, 2614 (1996)
\bibitem{SB2} Schumacher, B., Nielsen, M.A.: Quantum data processing and error correction. Phys. Rev. A \textbf{54}, 2629 (1996)
\bibitem{SLYS}Luo, S., Sun, Y.: Coherence and
complementarity in state-channel interaction. Phys. Rev. A.
\textbf{98}, 012113 (2018)
\bibitem{HF} Hansen, F.: Metric adjusted skew information. Proc. Natl. Aca. Sci. USA \textbf{105}, 9909
(2008)
\bibitem{EHL} Lieb, E. H.: Convex trace functions and the Wigner-Yanase-Dyson conjecture. Adv.
Math.\textbf{11}, 267 (1973)
\bibitem{BHATIA} Bhatia, R.: Partial traces and entropy inequalities. Linear Algebra Appl. \textbf{370} 125 (2003)




\bibitem{XKW} Xiong, C., Kumar, A., Wu, J.: Family of coherence measures and duality between quantum coherence and path distinguishability. Phys. Rev. A \textbf{98}, 032324 (2018)
\bibitem{BCD}Buscemi, F., Chiribella, G., D'Ariano, G.M.: Inverting quantum decoherence by classical feedback from the environment. Phys. Rev. Lett. \textbf{95}, 090501 (2005)
\bibitem{BHT}Br\'{a}dler, K., Hayden, P., Touchette, D., Wilde, M.M.: Trade-off capacities of the quantum Hadamard channels. Phys. Rev. A \textbf{81}, 062312 (2010)
\bibitem{STUM}Shitara, T., Ueda, M.: Determining the continuous family of quantum Fisher information from linear-response
theory. Phys. Rev. A \textbf{94}, 062316 (2016)
\bibitem{BS}Bagchi, S., Pati, A.K.: Uncertainty relations for general unitary operators. Phys. Rev. A \textbf{94},
042104 (2016)
\bibitem{THSN} Tajima, H., Shiraishi, N., Saito, K.: Uncertainty relations in implementation of unitary
operations. Phys. Rev. Lett. \textbf{121}, 110403 (2018)
\bibitem{BKTN}Bong, K.-W., Tischler, N., Patel, R.B., Wollmann, S., Pryde, G.J., Hall, M.J.W.: Strong unitary and
overlap uncertainty relations: theory and experiment. Phys. Rev.
Lett. \textbf{120}, 230402 (2018)
\bibitem{PAKSSU}Pati, A.K., Singh, U., Sinha, U.:
Measuring non-Hermitian operators via weak values. Phys. Rev. A
\textbf{92}, 052120 (2015)
\bibitem{NGS}Nirala, G., Sahoo, S.N., Pati, A.K., Sinha, U.: Measuring average of non-Hermitian operator
with weak value in a Mach-Zehnder interferometer. Phys. Rev. A
\textbf{99}, 022111 (2019)
\bibitem{ZXZC}Zhao, X., Zhang, C.: Uncertainty relations of non-Hermitian operators: theory and
experimental scheme. Front. Phys. \textbf{10}, 862868 (2022)


\vskip0.2in


\end{thebibliography}
\end{document}